\documentclass[aps,preprint,preprintnumber,nofootinbib,amsmath,amssymb,ascmac,bm,12pt]{revtex4}
\usepackage{bm}
\usepackage{graphicx}
\usepackage{dcolumn}
\usepackage{bm}

\usepackage{color}

\newcommand{\rd}{r_{\rm d}}
\newcommand{\rdo}{r_{\rm d1}}
\newcommand{\rdt}{r_{\rm d2}}

\newcommand{\thetah}{\hat{\theta}}
\newcommand{\phih}{\hat{\phi}}

\newcommand{\omi}{\omega_{\rm i}}
\newcommand{\omo}{\omega_{\rm o}}
\newcommand{\omp}{\omega_+}
\newcommand{\omm}{\omega_-}

\newcommand{\bi}{b_{\rm i}}
\newcommand{\bo}{b_{\rm o}}
\newcommand{\bpu}{b_+}
\newcommand{\bmi}{b_-}

\newcommand{\Rb}{R_{\rm b}}

\newcommand{\Ru}{R_{\rm u}}

\newcommand{\mz}{m_{(0)}}
\newcommand{\mo}{m_{(1)}}
\newcommand{\mt}{m_{(2)}}
\newcommand{\mth}{m_{(3)}}
\newcommand{\mf}{m_{(4)}}
\newcommand{\mc}{m_{\rm c}}

\newcommand{\sz}{\sigma_{(0)}}
\newcommand{\so}{\sigma_{(1)}}
\newcommand{\st}{\sigma_{(2)}}

\newcommand{\Go}{\varGamma_{(1)}}
\newcommand{\Gt}{\varGamma_{(2)}}

\newcommand{\Ron}{R_{(1)}}
\newcommand{\Rtw}{R_{(2)}}
\newcommand{\Rfo}{R_{(4)}}

\newcommand{\Uon}{U_{(1)}}

\newcommand{\Ufo}{U_{(4)}}

\newcommand{\Eze}{E_{(0)}}
\newcommand{\Eon}{E_{(1)}}
\newcommand{\Efo}{E_{(4)}}

\begin{document}

\title{On gravastar formation \\
{\it -- What can be the evidence of a black hole? --}}

\author{Ken-ichi Nakao$^1$}
\author{Chul-Moon Yoo$^2$}
\author{Tomohiro Harada$^3$}

\affiliation{
${}^{1}$
Department of Mathematics and Physics, Graduate School of Science, Osaka City University, 3-3-138 Sugimoto, Sumiyoshi, Osaka 558-8585, Japan\\
${}^{2}$Gravity and Particle Cosmology Group, Division of Particle and Astrophyisical Science, 
Graduate School of Science, Nagoya University, Nagoya 464-8602, Japan\\
${}^{3}$Department of Physics, Rikkyo University, Toshima, Tokyo 171-8501, Japan}


\begin{abstract}
Any observer outside black holes cannot detect any physical signal produced by 
the black holes themselves, since, by definition, the black holes are not located 
in the causal past of the outside observer. 
In fact, what we regard as black hole candidates in our view are not black holes but  will be 
gravitationally contracting objects. As well known, a black hole will form by 
a gravitationally collapsing object in the infinite future in the views of distant observers like us. 
At the very late stage of the gravitational collapse, the 
gravitationally contracting object behaves as a black body due to its gravity. 
Due to this behavior, the physical signals produced around it (e.g. the quasi-normal ringings 
and the shadow image) will be very similar to those caused in the eternal black hole spacetime. 
However those physical signals do not necessarily imply 
the formation of a black hole in the future, since we cannot rule out the possibility that the formation 
of the black hole is prevented by some unexpected event in the future yet unobserved. 
As such an example, we propose a scenario in which the final state of the gravitationally contracting 
spherical thin shell is a gravastar that has been proposed as a final configuration alternative to a black 
hole by Mazur and Mottola. This scenario implies that time necessary to observe the moment of the 
gravastar formation can be much longer than the lifetime of the present civilization, 
although such a scenario seems to be possible only if the 
dominant energy condition is largely violated. 
\end{abstract}

\preprint{OCU-PHYS-485}
\preprint{AP-GR-149}
\preprint{RUP-18-28}
\date{\today}
\maketitle

\section{Introduction}\label{sec:Intro}

The black hole is defined as a complement of the causal past of the 
future null infinity (see, e.g., \cite{Hawking,Wald}), or in physical terminology, 
a domain that is outside the view of any observer located outside it.  As well known, not only 
general relativity but also many of modified theories of gravity predict the formation of 
black holes through the gravitational collapse of massive objects in our universe. 
Many black hole candidates have been found through electromagnetic 
(see for example \cite{EMW}) and gravitational radiations\cite{GW}. 

\begin{figure}[!h]
\centering\includegraphics[width=5cm]{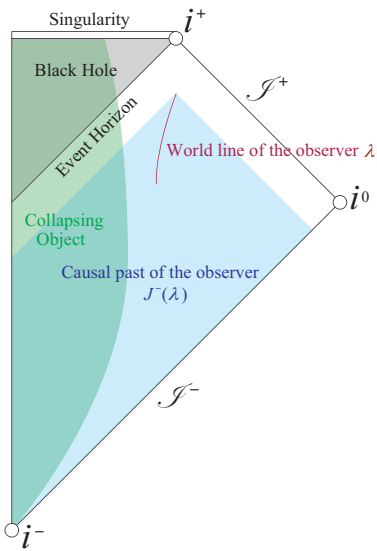}
\caption{ The conformal diagram of the black hole formation through the gravitational collapse of a 
spherical object is depicted. }
\label{collapsar}
\end{figure}

Hereafter any discussion in this paper will be basically based on general relativity. 
Figure \ref{collapsar} is the conformal diagram 
that describes the formation of a black hole through the gravitational collapse of 
a spherical massive object; the region shaded by gray is the black hole, the region 
shaded by green is the collapsing massive object,  the dark red curve $\lambda$ 
is the world line of a typical observer with finite lifetime outside the black hole  
and the region shaded by blue is the causal past of the observer often denoted by 
$J^-(\lambda)$: the causal past of the 
observer is defined as a set of all events which can be connected to 
$\lambda$ by causal curves, i.e., timelike or null curves; we believe that our situation in our universe 
is similar to the observer $\lambda$.  
Thus any event outside $J^-(\lambda)$ cannot causally affect the observer $\lambda$.  
As can be seen in Fig.~\ref{collapsar}, the black hole is outside the causal past of the observer $\lambda$, 
and hence  any signal detected by the observer $\lambda$    
(e.g., the black hole shadow, the quasi-normal ringing of gravitational 
radiation, the relativistic jet produced through Blandford-Znajek effect \cite{Blandford:1977}) 
cannot be caused by the back hole itself, although they strongly suggest 
the formation of the black hole as can be seen in Fig.~\ref{collapsar}; as well known, 
the black hole will form after infinite time has elapsed in the view of the distant observers. 

The black hole is often explained as an invisible astronomical object, but rigorously speaking, 
this explanation is inappropriate. We call an object invisible if it is in our view but 
does not emit anything detectable to our eyes or detectors. However, the black hole 
is located outside the view of the outside observer; 
this is the reason why the outside observer cannot see it.    
In the view of the outside observer, there is a gravitationally contracting object  
whose surface is asymptotically approaching the corresponding event horizon. 

Although the black hole is a promising final configuration of a gravitationally collapsing object in the 
framework of general relativity, various alternatives have been proposed 
 (see, for example, \cite{CP2017}; references are therein).  
We usually think that if a black hole candidate is not a black hole, it should be a static or stationary 
compact object, and believe that we will find differences from the 
black hole in observational data~\cite{CP2017}.
As mentioned in the above, any observer sees not black holes but gravitationally contracting objects and regard  
them as black holes. In the very late stage of the gravitational collapse of the massive object, 
distant observers can take a photo of the same shadow image as that in the eternal black hole spacetime 
with the boundary condition under which nothing is emitted from the white hole.  
Almost the same quasi-normal mode spectrum of gravitational waves as that 
of an eternal black hole spacetime will be generated around the contracting object in the very late stage 
and detected by distant observers. 
However, it should be noted that we cannot conclude from these observables that 
the black hole must form, since 
there is always the possibility that the formation of the black hole is prevented by some {\it unexpected 
events} and the contracting object settles down some alternative to the black hole in the future. 

\begin{figure}[!h]
\centering\includegraphics[width=10cm]{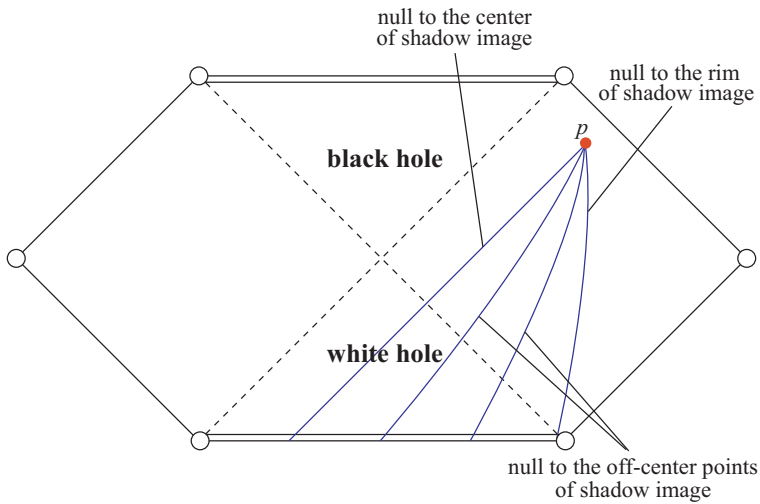}
\caption{ Typical null geodesics emanated from the event $p$ in past direction 
toward the shadow image observed at $p$ are depicted in the conformal diagram of the 
maximally extended Schwarzschild black spacetime. From this figure, we see that the shadow image 
is not produced by the absorption of photons into the black hole but is an image of the 
white hole with no radiation. If the white hole emits photons, the shadow image can be colored.  }
\label{shadow}
\end{figure}

In this paper, we revisit a very simple model which describes the gravitational collapse of 
an infinitesimally thin spherical shell and offer a scenario of the gravitational collapse  
accompanied by the formation of not a black hole but a gravastar that has been proposed as a 
final configuration of a gravitationally collapsing object alternative to a black hole 
by Mazur and Mottola~\cite{MM2004}. 
Our model shows that it is observationally very important when the gravastar formation begins. 
If the gravastar formation occurs in the very late stage of the gravitational collapse, 
the observers like as $\lambda$ will get shadow images and  
quasi-normal mode spectrum of gravitational waves which are almost the same as 
those of the maximally extended Schwarzschild spacetime 
with the boundary condition under which nothing 
emerges from the white hole\footnote{In the case of the maximally extended Schwarzschild spacetime, 
the so-called black hole shadow is the image of the white hole in the sense that if the white hole emits 
photons whose color is blue at distant observers, the shadow images are blue (see Fig.~\ref{shadow}).}. 
In this scenario, the {\it unexpected event} to prevent the black hole formation is the gravastar formation.   

This paper is organized as follows. In Sec.~II, we briefly review 
the basic equations to treat an infinitesimally thin spherical massive shell.  
In Sec.~III, based on the analyses of null rays in the spacetime with a 
spherical massive shell in Appendix A, we discuss why a massive object without the event horizon 
is regarded as a black hole candidate in the very late stage of its gravitational collapse in the view of 
distant observers. Then,  we give a model which represents a decay of the dust shell into two concentric 
timelike shells in Sec.~IV. In Sec.~V, we show a scenario in which 
a gravastar forms in the very late stage of the gravitational collapse of the dust shell; 
the gravastar formation is triggered by the decay of the dust shell. 
Sec.~VI is devoted to summary and discussion. 
In Appendix B, we show that the Bianchi identity leads to the conservation of the four momentum 
at the decay event.

In this paper, we adopt the abstract index notation, 
the sign conventions of the metric and Riemann tensors 
in Ref.~\cite{Wald} and basically the geometrized unit in which Newton's gravitational constant 
and the speed of light are one. If convenient, we adopt natural units with notice. 

\section{Equation of motion of a spherical shell}\label{sec:shell}

\begin{figure}[!h]
\centering\includegraphics[width=6cm]{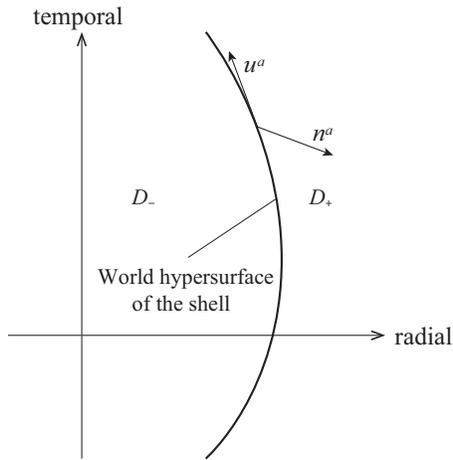}
\caption{ A schematic diagram of the situation considered in Sec.~II. The vertical direction is 
timelike, whereas the horizontal direction is spacelike. The world hypersurface 
of the shell is a bit thick curve. The four-velocity $u^a$ is the timelike unit tangent to and 
$n^a$ is the unit normal to the world hypersurface of the shell. }
\label{situation}
\end{figure}

In this section, we give basic equations to study the motion 
of a spherically symmetric massive shell which  
is infinitesimally thin and generates a timelike hypersurface
through its motion. We will refer this hypersurface as the world hypersurface of the shell. 
The world hypersurface of the shell divides the spacetime into two domains. 
These domains are denoted by $D_{+}$ and $D_{-}$.  
The situation is understood by Fig.~\ref{situation}. 

The geometry of the domains $D_{\pm}$ are assumed to be 
described by the Reissner-Nordstr\"{o}m-de Sitter spacetime; the infinitesimal world interval is given by 
\begin{eqnarray}
ds^2=-F_\pm (r)dt_\pm^2+\frac{1}{F_\pm(r)}dr^2
+r^2\left(d\theta^2+\sin^2\theta d\phi^2\right), \label{RN}
\end{eqnarray}
with
\begin{equation}
F_\pm(r)=1-\frac{2M_\pm}{r}+\frac{Q_\pm^2}{r^2}-\frac{\Lambda_\pm}{3}r^2,
\end{equation}
where $M_\pm$, $Q_\pm$ and $\Lambda_\pm$ are the mass parameter,  
the charge parameter and the cosmological constant in the domains $D_\pm$, respectively, 
whereas the gauge one-form in each domain is given by
\begin{equation}
A^\pm_\mu=\left(-\frac{Q_\pm}{r},0,0,0\right). 
\end{equation}
We should note that the time coordinate is 
not continuous at the shell and hence it is denoted by $t_+$ 
in the domain $D_+$ and by $t_-$ in the domain $D_-$, 
whereas $r$, $\theta$ and $\phi$ are everywhere continuous.

Since the finite energy and the finite momentum concentrate on the  
infinitesimally thin region, the stress-energy tensor diverges on the shell. This fact implies  
that the shell is categorized into the so-called scalar polynomial singularity \cite{Hawking-Ellis} 
through the Einstein equations. 
Even though the shell is a spacetime singularity, we can derive its equation of 
motion from the Einstein equations 
through Israel's formalism \cite{Israel1966}, since the singularity is so weak that its 
intrinsic metric on the world hypersurface of the shell exists 
and the extrinsic curvature defined on each side of the world hypersurface is finite.  
Hence, hereafter, we do not regard the shell as  
a spacetime singularity. 

We cover the neighborhood of the world hypersurface of the shell by the Gaussian normal coordinate $\lambda$, 
where $\partial/\partial\lambda$ is a 
unit vector normal to the shell and directs from 
$D_-$ to $D_+$. Then, the sufficient condition 
to apply Israel's formalism is that the stress-energy tensor is written in the form
$$
T^{ab}=S^{ab}\delta(\lambda-\lambda_{\rm s}),
$$
where the shell is located at $\lambda=\lambda_{\rm s}$, $\delta(x)$ is Dirac's delta function, 
and $S^{ab}$ is the surface stress-energy tensor on the shell. 

We impose that the metric tensor $g_{ab}$ is continuous even at the shell. 
Hereafter, $n^a$ denotes the unit normal vector to the shell, 
instead of $\partial/\partial \lambda$. 
The intrinsic metric of the world hypersurface of the shell is given by
$$
h_{ab}=g_{ab}-n_a n_b,
$$
and the extrinsic curvature is defined as
$$
K^{\pm}_{ab}=-h^{c}{}_a h^{d}{}_b \nabla^{(\pm)}_c n_d,
$$
where $\nabla^{(\pm)}_c$ is the covariant derivative with respect to the metric in the 
domain $D_\pm$. This extrinsic curvature describes how the world hypersurface of the shell 
is embedded into the domain $D_\pm$. In accordance with Israel's formalism, the Einstein equations lead to
\begin{equation}
K^{+}_{ab}-K^{-}_{ab}=8\pi\left(S_{ab}-\frac{1}{2}h_{ab}{\rm tr}S\right),
\label{j-condition}
\end{equation}
where ${\rm tr}S$ is the trace of $S_{ab}$.  Equation (\ref{j-condition}) gives us the condition 
of the metric junction. 

By the spherical symmetry of the system, the surface stress-energy tensor of the shell 
should be the perfect fluid type; 
$$
S_{ab}=\sigma u_a u_b+P\left(h_{ab}+u_a u_b\right),
$$
where $\sigma$, $P$ and $u^a$ are the energy per unit area, the tangential pressure and the four-velocity, respectively. 
Due to the spherical symmetry of the system, 
the motion of the shell is described in the form of $t_\pm=T_\pm(\tau)$ and 
$r=R(\tau)$, where $\tau$ is the proper time of the shell.  
The 4-velocity is given by 
$$
u^\mu=\left(\dot{T}_{\pm},\dot{R},0,0\right),
$$
where a dot represents a derivative with respect to $\tau$. 
Then, $n_\mu$ is given by
$$
n_\mu=\left(-\dot{R},\dot{T}_{\pm},0,0\right).
$$
Together with $u^\mu$ and $n^\mu$, the following unit vectors form an orthonormal frame;
\begin{align}
\thetah^\mu&=\left(0,0,\frac{1}{r},0\right), \cr
\phih^\mu&=\left(0,0,0,\frac{1}{r\sin\theta}\right). \nonumber
\end{align}

The extrinsic curvature is obtained as
\begin{align}
K_{ab}^\pm u^a u^b&=\frac{1}{F_{\pm}\dot{T}_{\pm}}\left(\ddot{R}+\frac{F'_\pm}{2}\right), \cr
K^\pm_{ab}\thetah^a \thetah^b&=
K_{ab}^\pm\phih^a \phih^b=-n^a\partial_a \ln r|_{D_\pm}=-\frac{F_\pm}{R}\dot{T}_\pm  \label{K-thth}
\end{align}
and the other components vanish, where 
$$
F_\pm=F_\pm(R),
$$ 
and a prime represents a derivative with respect to its argument, i.e., 
$$
F'_\pm=\frac{dF_\pm(R)}{dR}.
$$
By the normalization condition $u^\mu u_\mu=-1$, we have
\begin{equation}
\dot{T}_{\pm}= \frac{1}{F_\pm}\sqrt{\dot{R}^2+F_\pm}~, \label{t-dot}
\end{equation}
where we have assumed that the shell exists outside the black hole and $u^a$ is future-directed. 
Substituting Eq.~(\ref{t-dot}) into Eq.~(\ref{K-thth}), we have
\begin{equation}
K_{ab}^\pm\thetah^a \thetah^b=- \frac{1}{R}\sqrt{\dot{R}^2+F_\pm}. \label{K-thth-2}
\end{equation}
From Eqs.~(\ref{j-condition}) and (\ref{K-thth-2}), we have
\begin{equation}
- \frac{1}{R}\sqrt{\dot{R}^2+F_+}+ \frac{1}{R}\sqrt{\dot{R}^2+F_-}=4\pi \sigma. 
\label{thth-component}
\end{equation}
Hereafter, we assume the weak energy condition $\sigma\geq0$. Then, Eq.~(\ref{thth-component}) 
leads to
\begin{equation}
F_->F_+. \label{M-increasing}
\end{equation}

From the $u$-$u$ component of Eq.~(\ref{j-condition}), we obtain the following relations.
\begin{equation}
\frac{d m}{d\tau}+4\pi P\frac{dR^2}{d\tau}=0, \label{E-con}
\end{equation}
where $m$ is the proper mass of the shell defined as
\begin{equation}
m:=4\pi \sigma R^2. \label{p-mass}
\end{equation}
By dividing both sides of Eq.~(\ref{E-con}) by $dR/d\tau$, we have
\begin{equation}
\frac{dm}{dR}+8\pi PR=0. \label{E-con2}
\end{equation}
By giving the equation of state to determine $P$, 
Eq.~(\ref{E-con2}) determines the dependence of $m$ 
on $R$. Equation (\ref{E-con}) implies that if the shell is composed of the dust, i.e., 
$P=0$, $m$ is constant. 
Hereafter, we assume $\sigma$ is positive and hence $m$ is also positive. 

In general, the energy cannot be uniquely defined within the framework of general relativity. 
However, in the case of the spherically symmetric spacetime, quasi-local energies proposed by many 
researchers agree with the so-called Misner-Sharp energy (see for example Ref.~\cite{Hayward}). 
The Misner-Sharp energy just on each side of the shell is given as
$$
{\cal M}_\pm=\frac{R}{2}\left(1-F_\pm\right).
$$
Hence, the Misner-Sharp energy included by the shell is given by
\begin{equation}
{\cal M}=\frac{R}{2}\left(F_--F_+\right). \label{MS-energy}
\end{equation}

From Eq.~(\ref{thth-component}), we have
\begin{equation}
\sqrt{\dot{R}^2+F_\pm}\pm\frac{m}{R}=\sqrt{\dot{R}^2+F_\mp}, \label{thth-component-2}
\end{equation}
where we have used Eq.~(\ref{p-mass}). 
By taking the square of Eq.~(\ref{thth-component-2}), we obtain
\begin{equation}
\sqrt{\dot{R}^2+F_\pm}=E\mp\frac{m}{2R},
\label{thth-component-3}
\end{equation}
where 
\begin{equation}
E=\frac{R}{2m}\left(F_--F_+\right)=\frac{\cal M}{m}
\label{E-def}
\end{equation}
is the specific energy of the shell. 
By taking the square of Eq.~(\ref{thth-component-3}), we obtain the energy equations for the shell as follows;
\begin{equation}
\dot{R}^2+U(R)=0 \label{E-equation}
\end{equation}
with 
\begin{equation}
U(R)=F_\pm-\left(E\mp\frac{m}{2 R}\right)^2.
\label{potential}
\end{equation}
Here note that unless $P=0$, $m$ and $E$ in Eq.~(\ref{potential}) depend on $R$. 

Since the left-hand side of Eq.~(\ref{thth-component-3}) is positive, the right-hand side should also be positive; 
\begin{equation}
E-\frac{m}{2R}>0.
\label{future-directed}
\end{equation}
By substituting Eq.~(\ref{thth-component-3}) into Eq.~(\ref{t-dot}), we have 
\begin{equation}
\dot{T}_\pm=\frac{1}{F_\pm}\left(E\mp\frac{m}{2R}
\right). \label{t-dot-2}
\end{equation}
Here note again that Eq.~(\ref{thth-component}) is obtained under the assumption that 
the shell is located outside the black hole. If the shell is in the black hole, Eq.~(\ref{future-directed}) 
is not necessarily satisfied, and accordingly, $\dot{T}_\pm$ is not necessarily positive. 

\section{The very late stage of the gravitationally contracting shell}

In Appendix A, by studying null rays in the spacetime with a 
spherical shell, we show that the {\it contracting} shell with the radius 
very close to its gravitational radius effectively behaves as a black body due to its gravity, 
even though the material of the shell causes the specular reflection of or is transparent to null rays: 
both the null ray reflected by the shell and that transmitted through the shell suffer the large redshift 
or are trapped in the neighborhood of the shell.  
Hence the behavior of any physical field in this spacetime  
will be very similar to those in the maximally extended  
Schwarzschild spacetime with the boundary condition 
under which nothing appears from the white hole: the contracting 
shell corresponds to the white hole horizon. 
In the late stage of the gravitational collapse, the image of the shell 
and the spectrum of the quasi-normal modes will be very similar to 
the black hole shadow and the quasi-normal modes of the 
Schwarzschild spacetime. 
By contrast, the {\it static} shell will show images distinctive from the black hole shadow and 
a  quasi-normal mode spectrum, of the Schwarzschild spacetime, 
since it does not behave as a black body.

\section{Decay of a timelike shell; conservation law}

In this section, we consider the decay process of a spherical massive shell into two 
daughter spherical shells concentric with the parent shell; in the next section, 
this decay process is regarded as a trigger of the gravastar formation. 

\begin{figure}[htbp]
 \begin{center}
 \includegraphics[width=7cm,clip]{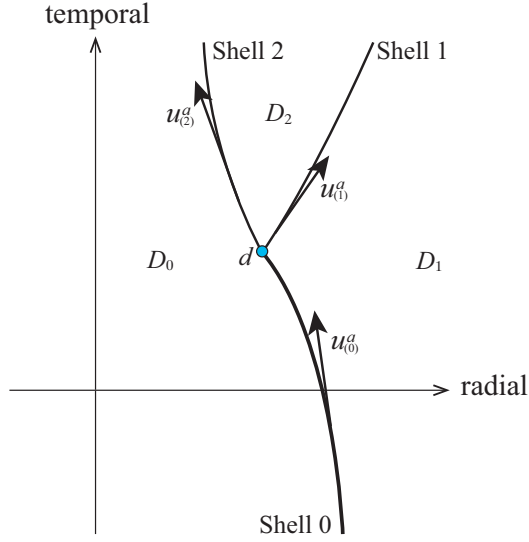}
 \end{center}
 \caption{
The schematic diagram representing the decay of Shell 0 into Shell 1 and Shell 2 is depicted.   }
 \label{Situation2}
\end{figure}

We call the parent shell Shell~0 and assume that Shell~0 initially contracts but 
decays just before the formation of a black hole.  
One of two daughter shells called Shell 1 is located outside the other one called Shell 2 
(see Fig.~\ref{Situation2}). 
Shell 0, Shell 1 and Shell 2 divide the spacetime into three domains: $D_0$ is the domain 
whose boundary is composed of Shell 0 and Shell 2, $D_1$ is the domain 
whose boundary is composed of Shell 0 and Shell 1, and $D_2$ is the 
domain whose boundary is composed of Shell 1 and Shell 2.   

The infinitesimal world intervals of the three domains $D_i$ ($i=0,1,2$) are given as
$$
ds^2=-F_i(r)dt_i^2+\frac{dr^2}{F_i(r)}+r^2(d\theta^2+\sin^2\theta d\phi^2),
$$
where
$$
F_i(r)=1-\frac{2M_i}{r}+\frac{Q_i^2}{r^2}-\frac{\Lambda_i}{3} r^2.
$$

For later convenience, we introduce the dyad basis related to the two-sphere whose components 
are, in all domain, given as  
\begin{align}
\thetah^\mu&=\left(0,0,\frac{1}{r},0\right), \\
\phih^\mu&=\left(0,0,0,\frac{1}{r\sin\theta}\right).
\end{align}
By virtue of the spherical symmetry, the surface stress-energy tensor of Shell $I$ ($I=0,1,2$) is given in the form,
\begin{equation}
S^{ab}_{(I)}=\sigma_{(I)}u_{(I)}^a u_{(I)}^b+P_{(I)}H^{ab},
\end{equation}
where $\sigma_{(I)}$, $P_{(I)}$ and $u_{(I)}^a$ are the surface energy density, the tangential pressure 
and the four-velocity of Shell $I$, respectively, and 
$$
H^{ab}=\thetah^a \thetah^b+ \phih^a \phih^b.
$$
We assume that $\sigma_{(I)}$ is positive. 

The radial coordinate of the decay event $d$ is denoted by $r=\rd$. 
Hereafter, the time and radial coordinates of Shell $I$ are denoted by $T_{(I)i}$ and $R_{(I)}$, 
where $i$ is the index to specify the time coordinate in the domain $D_i$ ($i=0,1,2$): as mentioned, 
the time coordinate is not continuous at the shells. 
Then, we introduce the orthonormal basis of the center of mass frame at $d$: 
the components of them are given as
\begin{align}
u^\alpha_{(0)i}&=\left(\dot{T}_{(0)i},\dot{R}_{(0)},0,0\right), \label{u-def}\\
n^\alpha_{(0)i}&=\left(\frac{\dot{R}_{(0)}}{F_i(\rd)},F_i(\rd)\dot{T}_{(0)i},0,0\right), \\
\thetah^\alpha&=\left(0,0,\frac{1}{\rd},0\right), \\
\phih^\alpha&=\left(0,0,0,\frac{1}{\rd\sin\theta}\right), \label{phi-def}
\end{align}
where $i=0$ ($i=1$) represents the components in $D_0$ ($D_1$), 
and a dot means the derivative with respect to the proper time of Shell 0. 

Hereafter, we assume that the decay occurs before Shell 0 forms a black hole, i.e., 
\begin{equation}
F_2(\rd)>0. \label{rd-cons}
\end{equation}

The four-velocity $u_{(J)}^a$ ($J=1,2$) at $d$ is written in the form, 
\begin{equation}
u_{(J)}^a=\varGamma_{(J)} u_{(0)}^a+\epsilon_{(J)}\sqrt{\varGamma_{(J)}^2-1} ~n_{(0)}^a, \label{uj-comp}
\end{equation}
where $\varGamma_{(J)}$ is a positive number larger than one, and 
$\epsilon_{(J)}=\pm1$ is the sign factor which will be fixed by the momentum conservation. 

We require the conservation of four-momentum at $d$; 
\begin{equation}
\mz u_{(0)}^a=\mo u_{(1)}^a+\mt u_{(2)}^a, \label{MC-2}
\end{equation} 
where 
$$
m_{(I)}:=4\pi R_{(I)}^2 \sigma_{(I)}. 
$$
Note that $m_{(I)}$ is positive since we assume $\sigma_{(I)}$ is positive.  
The derivation of the conservation law from the Bianchi identity is shown in Appendix B. 

The $u$-component of Eq.~(\ref{MC-2}) leads to
\begin{equation}
\mz=\mo \varGamma_{(1)}+\mt\varGamma_{(2)}, \label{MC-u}
\end{equation}
whereas the $n$-component leads to
\begin{equation}
0=\mo\epsilon_{(1)}\sqrt{\varGamma_{(1)}^2-1}+\mt\epsilon_{(2)}\sqrt{\varGamma_{(2)}^2-1}~. \label{MC-n}
\end{equation}
Since $m_{(J)}$ is positive, Eq.~(\ref{MC-n}) implies that $\epsilon_{(1)}=+1$ and $\epsilon_{(2)}=-1$ should hold 
in the situation we consider. 

From Eq.~(\ref{MC-u}), we have
\begin{equation}
\mt^2\Gt^2=\mz^2-2\mz\mo\Go+\mo^2\Go^2, \label{gamma2-1}
\end{equation}
whereas, from Eq.~(\ref{MC-n}), we have
\begin{equation}
\mt^2\Gt^2=\mo^2\left(\Go^2-1\right)+\mt^2. \label{gamma2-2}
\end{equation}
Equations (\ref{gamma2-1}) and (\ref{gamma2-2}) lead to 
\begin{equation}
\Go=\frac{\mz^2+\mo^2-\mt^2}{2\mz\mo}. \label{gamma1}
\end{equation}
Through the similar procedure, we obtain
\begin{equation}
\Gt=\frac{\mz^2+\mt^2-\mo^2}{2\mz\mt}. \label{gamma2}
\end{equation}
Equations (\ref{gamma1}) and (\ref{gamma2}) lead to
\begin{equation}
\mz=\mo\Go+\mt\Gt \geq \mo+\mt. \label{mass-order}
\end{equation}

From Eq.~(\ref{uj-comp}) with $J=1$, we have
\begin{align}
u_{(1)1}^t&=\Go u_{(0)1}^t+\sqrt{\Go^2-1}~n_{(0)1}^t \nonumber \\
&=\frac{\Go}{F_1}\left[\frac{\rd}{2\mz}\left(F_0-F_1\right)-\frac{\mz}{2\rd}\right]+\sqrt{\Go^2-1}\frac{\dot{R}_{(0)}}{F_1}, 
\label{ut1-1}
\end{align}
where we have used Eqs.~(\ref{E-def}) and (\ref{t-dot-2}) for $u_{(0)1}^t$, and
$
F_i=F_i(\rd).
$
On the other hand, by using Eq.~(\ref{t-dot-2}), we have
\begin{equation}
u_{(1)1}^t=\frac{1}{F_1}\left[\frac{\rd}{2\mo}\left(F_2-F_1\right)-\frac{\mo}{2\rd}\right]. \label{ut1-2}
\end{equation}
Then Eqs.~(\ref{ut1-1}) and (\ref{ut1-2}) lead to
\begin{equation}
F_2=F_1+\frac{\mo^2}{\rd^2}+\frac{2\mo}{\rd}
\left\{\Go\left[\frac{\rd}{2\mz}\left(F_0-F_1\right)-\frac{\mz}{2\rd}\right]+\sqrt{\Go^2-1} \dot{R}_{(0)}\right\}. \label{F2-1}
\end{equation}
By the similar procedure starting from Eq.~(\ref{uj-comp}) with $J=2$, we have
\begin{equation}
F_2=F_0+\frac{\mt^2}{\rd^2}-\frac{2\mt}{\rd}
\left\{\Gt\left[\frac{\rd}{2\mz}\left(F_0-F_1\right)+\frac{\mz}{2\rd}\right]-\sqrt{\Gt^2-1} \dot{R}_{(0)}\right\}. \label{F2-2}
\end{equation}
By using Eqs.~(\ref{gamma1}) and (\ref{gamma2}), we can see that Eq.~(\ref{F2-1}) is equivalent to Eq.~(\ref{F2-2}). 
The momentum conservation (\ref{MC-2}) uniquely determines the geometry of $D_2$ which 
appears after the decay event $d$ if we fix the values of $\rd$, $F_0$, $F_1$, $\mz$, $\mo$ and $\mt$; 
$\dot{R}_{(0)}$ is determined through Eqs.~(\ref{E-equation}) and (\ref{potential}) except for its sign 
that we have to choose.

\section{Gravastar formation}

In this section, we consider the gravitational collapse of Shell 0 accompanied by 
the gravastar formation. Here, we will adopt the gravastar model devised  
by Visser and Wiltshire (VW)~\cite{VW2004}, 
which is simpler and clearer than the original one of Mazur and E.~Mottola; VW gravastar is 
a spherical de Sitter domain surrounded by a spherical infinitesimally thin shell. 

We assume that Shell 0 is an electrically neutral dust shell, $P_{(0)}=0$; 
the geometry of its inside is Minkowskian, whereas 
that of its outside is Schwarzschildian; $M_0=Q_0=Q_1=0=\Lambda_0=\Lambda_1$ hold. 
From Eq.~(\ref{potential}), we obtain the effective potential of Shell 0 as
$$
U_{(0)}\left(R_{(0)}\right)=1-\left(E_{(0)}+\frac{\mz}{2R_{(0)}}\right)^2,
$$
where $\mz$ is constant due to the conservation law (\ref{E-con2}), 
and hence $E_{(0)}=M_1/\mz$ is also constant. 

From Eq.~(\ref{future-directed}), we have 
\begin{equation}
R_{(0)}>\frac{M_1}{2E_{(0)}^2} \label{R0-cons}
\end{equation}
so that $u_{(0)}^a$ is future-directed for $R_{(0)}>2M_1$. We are interested in the case that  
Shell 0 contracts and forms a black hole, i.e., $R_{(0)}\leq2M_1$, if the decay of Shell 0 
does not occur. Hence we assume  
\begin{equation}
E_{(0)}>\frac{1}{2} \label{E0-cons}
\end{equation}
so that the r.h.s. of Eq.~(\ref{R0-cons}) is less that $2M_1$. 
Then, by investigating the effective potential $U_{(0)}$, 
we can easily see that the allowed domain for the motion of Shell 0 is 
$$
0\leq R_{(0)} \leq \frac{M_1}{2E_{(0)}(1-E_{(0)})},
$$
for $1/2<E_{(0)}<1$, whereas 
$$
0\leq R_{(0)}<\infty
$$
for $E_{(0)}\geq1$. 

\begin{figure}[htbp]
 \begin{center}
 \includegraphics[width=8cm,clip]{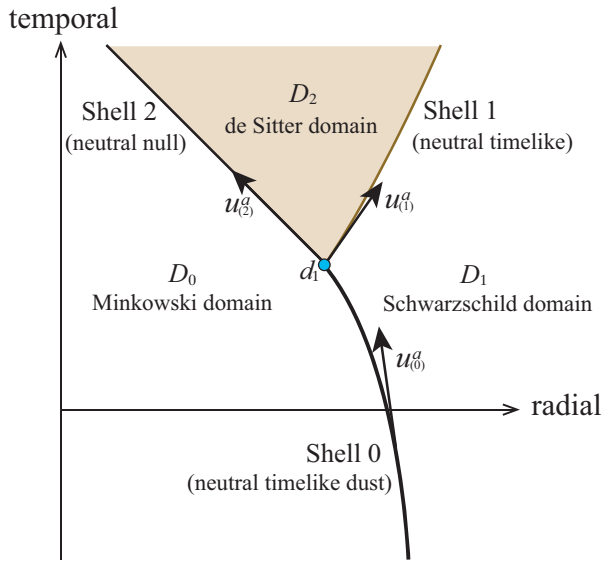}
 \end{center}
 \caption{
The schematic diagram representing the formation of a gravastar triggered by 
decay of Shell 0 into Shell 1 and Shell 2 is depicted.   }
 \label{G-formation}
\end{figure}

We assume that the formation of the gravastar is triggered by the decay 
of Shell~0 into Shell~1 and Shell~2. The domain $D_2$ between Shell~1 and Shell~2 is described 
by the de Sitter geometry, i.e., $M_2=Q_2=0$ but $\Lambda_2>0$. Shell~2 shrinks to zero radius, 
so that the innermost domain $D_0$ disappears at some stage (see Fig.~\ref{G-formation}). 
By contrast, Shell 1 corresponds to the crust of the gravastar. 
The decay event of Shell 0 and its areal radius are denoted by $d_1$ and $\rdo$, respectively. 

It is observationally very important when the gravaster formation starts. 
In Ref.~\cite{MM2004}, the gravastar formation is implicitly assumed to 
start when the radius $R$ of the contracting object satisfies $R-2M\simeq l_{\rm pl}$, where $l_{\rm pl}$ ($\simeq 1.6\times10^{-33}$cm) is the Planck length.  
The time scale in which the radius of the collapsing object satisfiies  $0<R-2M\ll 2M$ is almost equal to the free fall time of the system. 
From Eqs.~(\ref{E-equation}), (\ref{potential}) and (\ref{t-dot-2}), we can see that 
once $0<R-2M\ll 2M$ is satisfied, the time evolution of the radius of a dust shell ($m$=constant) is given by $R\simeq {\rm Const.}\times e^{-\frac{t}{2M}}$, 
where $t$ is the proper time for an asymptotic observer. 
Thus, the time scale in which $R-2M\simeq l_{\rm pl}$ is achieved  
will be much less than our average lifetime if the mass $M$ of the contracting object takes $1M_\odot <M<10^{8}M_\odot$, 
where $M_\odot$ is the solar mass. 
If the criterion of the gravastar formation proposed by Mazur and Mottola is correct, 
we can, in principle, observe the gravastar as a final product of the gravitational collapse of a massive object.  
However, we know no physically well motivated estimate on when the gravastar formation starts. There is the possibility 
that the gravastar formation may start at very late stage of the gravitational collapse. 
For example, the trigger of the gravastar formation might be the energy loss from the system 
due to the semi-classical effects associated to the gravitational collapse. 
If the contracting object has a mass larger than the solar mass $M_\odot$, the particles created through the semi-classical effect will be photons and gravitons; for a 
spherically symmetric contracting object with $R-2M \ll 2M$, the time variation of the mass of the contracting object will be governed by
\begin{equation}
\frac{dM}{dt}= -\frac{\pi^2}{30} T_{\rm BH}^4 A_{\rm H}, \label{H-flux}
\end{equation}
in natural units, where, $m_{\rm pl}$ being the Planck mass, $T_{\rm BH}=m_{\rm pl}^2/8\pi M$ is the Bekenstein-Hawking temperature, 
and $A_{\rm H}$ is the horizon area $16\pi M^2/m_{\rm pl}^4$\cite{PP}. We assume that after a small fraction $\epsilon$ ($\ll1$) of the initial mass of the collapsing 
objects is released through the particles created by the semi-classical effect, the gravastar formation begins. Then by solving Eq.~(\ref{H-flux}), we can see that 
the time scale $t_\epsilon$ in which the initial mass $M_{\rm i}$ of the 
contracting object becomes $(1-\epsilon)M_{\rm i}$ is given by$$
t_\epsilon=7680\pi \left[1+{\cal O}(\epsilon)\right]\epsilon\frac{M_{\rm i}^3}{m_{\rm pl}^4}=3.1\times 10^{67} \left[1+{\cal O}(\epsilon)\right]\epsilon\left(\frac{M_{\rm i}}{M_\odot}\right)^3 ~{\rm yr}.
$$
If this is true, asymptotic observers should wait to observe the gravastar formation for very long time after 
the gravitational collapse has begun: the time will be much longer than the age of the universe for a 
black hole of the mass larger than the solar mass if $\epsilon \gtrsim 10^{-50}$.

Anyway, the radius of Shell~0 might be very close 
to the gravitational radius in the domain $D_1$ when the gravastar formation starts. Hence hereafter 
we assume so.

\subsection{The motion of Shell 2}

Let us start on the discussion about Shell 2. We assume that Shell 2 moves inward 
with the energy much larger than its proper mass, i.e., $E_{(2)}={\cal M}_{(2)}/m_{(2)}\gg1$, 
where ${\cal M}_{(2)}$ is the Misner-Sharp energy of Shell 2. 
We introduce 
\begin{equation}
k_{(2)}^a:=\frac{\mt u_{(2)}^a}{{\cal M}_{(2)}}
\end{equation}
and rewrite $S_{(2)}^{ab}$ in the form
\begin{align}
S_{(2)}^{ab}=E_{(2)}\left[\frac{{\cal M}_{(2)}}{4\pi R_{(2)}^2}k_{(2)}^a k_{(2)}^b +\frac{\mt}{{\cal M}_{(2)}}H^{ab}P_{(2)}\right]. 
\end{align}
We assume that $\sigma_{(2)}$ is non-negative, and the equation of state is given by  
$$
P_{(2)}=w_{(2)}\sigma_{(2)},
$$ 
where $\sigma_{(2)}$ $w_{(2)}$ is a constant number of $\left|w_{(2)}\right|\leq1$. 
Then we take the massless limit for Shell 2: $\mt\rightarrow 0$ with the Misner-Sharp energy ${\cal M}_{(2)}$ 
fixed. From Eqs.~(\ref{uj-comp}) and (\ref{gamma2}), we have 
\begin{equation}
k_{(2)}^a\longrightarrow \frac{\mz^2-\mo^2}{2\mz{\cal M}_{(2)}}\left(u_{(0)}^a-n_{(0)}^a\right).
\end{equation}
It is easy to see that $k_{(2)}^a$ is null in this limit. Furthermore, we have
\begin{equation}
\frac{S_{(2)}^{ab}}{E_{(2)}}\longrightarrow \frac{{\cal M}_{(2)}}{4\pi R_{(2)}^2}k_{(2)}^a k_{(2)}^b.  
\end{equation}
As expected, Shell 2 becomes the null dust in this limit. 
Although $S^{ab}_{(2)}$ itself diverges due to the Lorentz contraction, the Misner-Sharp mass 
kept by Shell 1 is finite by assumption (see Eq.~(\ref{MS-energy})): this divergence should be absorbed 
in the integral measure (please see Ref.~\cite{Poisson} for the proper stress-energy tensor of the null shell). 
In the massless limit of Shell~2, we have ${\cal M}_{(2)}$ at the decay event, from Eq.~(\ref{F2-2}), in   
the following form;
\begin{align}
\left. {\cal M}_{(2)}\right|_{d_1}= \frac{\mz^2-\mo^2}{2\mz}\left[\Eze+\frac{\mz}{2\rdo}
+\sqrt{\left(\Eze-\frac{\mz}{2\rdo}\right)^2-F_1}\right]. \label{calM2}
\end{align}

The cosmological constant in $D_2$ is determined at the decay event $d_1$ through 
\begin{equation}
\Lambda_2=\frac{6}{\rdo^3}\left.{\cal M}_{(2)}\right|_{d_1}. \label{Lambda}
\end{equation}
Then the Misner-Sharp energy of Shell 2 
is a function of the radius of Shell 2; 
\begin{equation}
{\cal M}_{(2)}=\frac{\Rtw}{2}(F_0-F_2)=\frac{\Lambda_2}{6}\Rtw^3. \label{calM2-2}
\end{equation}
As can be seen from Eq.~(\ref{calM2-2}), ${\cal M}_{(2)}$ vanishes  
when the radius of Shell 2 becomes zero, or in other words, Shell 2 disappears when it shrinks 
to the symmetry center $r=0$.  

\subsection{The motion of Shell 1}

From Eq.~(\ref{uj-comp}), the radial velocity of Shell 1 at $d_1$ is written in the form
\begin{align}
\dot{R}_{(1)}&=\Go \dot{R}_{(0)}+\sqrt{\Go^2-1}F_1\dot{T}_{(0)1} \nonumber \\
&=-\Go \sqrt{\left(\Eze-\frac{\mz}{2\rdo}\right)^2-F_1} +\sqrt{\Go^2-1}\left(\Eze-\frac{\mz}{2\rdo}\right).
\end{align}
Hence, $\dot{R}_{(1)}$ is positive at $d_1$, if and only if 
\begin{equation}
\Go^2 > \frac{1}{F_1}\left(\Eze-\frac{\mz}{2\rdo}\right)^2 \label{gamma1-exp}
\end{equation}
holds. 
Taking into account Eqs.~(\ref{gamma1}) and (\ref{mass-order}), Eq.~(\ref{gamma1-exp}) leads to the condition on $\mo$ as
\begin{equation}
\left.\mo\right|_{d_1}<\mc:=\frac{\mz}{\sqrt{F_1}}\left[\Eze-\frac{\mz}{2\rdo}-\sqrt{\left(\Eze-\frac{\mz}{2\rdo}\right)^2-F_1}\right].
\label{m1-exp}
\end{equation}
On the other hand, $\dot{R}_{(1)}$ is negative or zero, if and only if 
\begin{equation}
\Go^2 \leq \frac{1}{F_1}\left(\Eze-\frac{\mz}{2\rdo}\right)^2 \label{gamma1-con}
\end{equation}
or equivalently, 
\begin{equation}
\mc \leq\mo<\mz
\label{m1-con}
\end{equation}
at $d_1$. 

The effective potential $\Uon$ of Shell 1 is given by
$$
\Uon\left(\Ron\right)=1-\frac{2M_1}{\Ron}
-\left[\frac{1}{\mo}\left(M_1-\frac{\Lambda_2}{6}\Ron^3\right)-\frac{\mo}{2\Ron}\right]^2.
$$
The future directed condition $\dot{T}_{(1)1}>0$ implies
\begin{equation}
\frac{1}{\mo}\left(M_1-\frac{\Lambda_2}{6}\Ron^3\right)-\frac{\mo}{2\Ron}>0. \label{future-g}
\end{equation}
It is easy to see that,  irrespective of the equation of state of Shell~1,  
\begin{equation}
\Ron<\Ru:=\left(\frac{6M_1}{\Lambda_2}\right)^{1\over3}  \label{Ru}
\end{equation}
is necessary so that Eq.~(\ref{future-g}) is satisfied, since we require 
$\sigma_{(1)}\geq0$, or equivalently, $\mo\geq0$; 
the allowed domain for the motion of Shell~1 is bounded from above.

\subsubsection{Dissipation through further decay}

\begin{figure}[htbp]
 \begin{center}
 \includegraphics[width=8cm,clip]{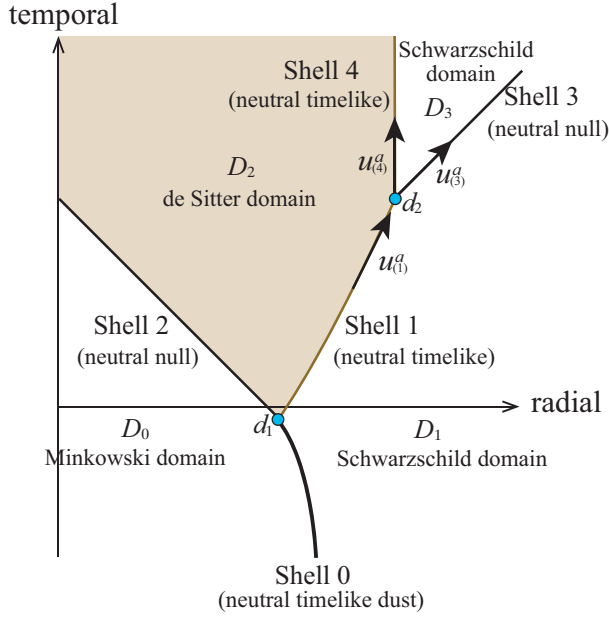}
 \end{center}
 \caption{
The schematic diagram representing the stabilization of the gravastar 
due to the decay of Shell 1 into Shell 3 and Shell 4. Shell 3 is null and causes the 
dissipation which results in the stabilization of the crust of the gravastar, i.e., Shell 3.  }
 \label{Emission}
\end{figure}

Shell 1 is the crust of the gravastar. It is dynamical and hence  
should dissipate its energy so that the gravastar is stable and static. 
Chan et al studied the gravastar formation by taking into account a dissipation through the 
emission of null dust\cite{CSRW}. 
In this paper, instead of the emission of the null dust, we 
assume that the crust, Shell 1, emits outward Shell 3 at the event $d_2$ with $r=\rdt$ and becomes static and stable; 
the static crust of the gravastar is called Shell 4.  
This process is equivalent to the decay of Shell 1 into Shell 3 and Shell 4  (see Fig.~\ref{Emission}). 

The domain between Shell~3 and Shell~4 is denoted by $D_3$. 
Replacing Shell~0, Shell~1, Shell~2, $D_0$ and $D_2$ 
by Shell~1, Shell~3, Shell~4, $D_2$ and $D_3$  in Eq.~(\ref{F2-1}), 
the same argument as that in Sec.~IV 
is applied, and we obtain
\begin{equation}
F_3=F_1
+\frac{\mth^2}{\rdt^2}+\frac{2\mth}{\rdt}\left\{\varGamma_{(3)}\left[\frac{\rdt}{2\mo}
\left(F_2-F_1\right)
-\frac{\mo}{2\rdt}\right]+\sqrt{\varGamma_{(3)}^2-1}\dot{R}_{(1)}\right\}, \label{F3}
\end{equation}
where 
$$
\varGamma_{(3)}=\frac{\mo^2+\mth^2-\mf^2}{2\mo\mth}, 
$$
and, in this section~VB{\it 1}, all quantities are evaluated  at $d_2$. 
Here, as in the case of Shell 2, we take the limit $\mth\rightarrow0$ 
under the assumption of $P_{(3)}=w_{(3)}\sigma_{(3)}$ with $w_{(3)}$ fixed: this limit is 
equivalent to the assumption that Shell 3 is a null dust. From Eq.~(\ref{F3}), we have
$$
F_3 =F_1+\frac{\mo^2-\mf^2}{\mo\rdt}\left(\Eon-\frac{\mo}{2\rdt}+\dot{R}_{(1)}\right).
$$
Then, from this result, we have
\begin{align}
\Efo&=\frac{\rdt}{2\mf}\left(F_2-F_3\right) \nonumber \\
&=\frac{\rdt}{2\mf}\left(F_2-F_1\right)+\frac{\rdt}{2\mf}\left(F_1-F_3\right) \nonumber \\
&=\frac{\mo}{\mf}\Eon-\frac{\mo^2-\mf^2}{2\mo\mf}\left(\Eon-\frac{\mo}{2\rdt}+\dot{R}_{(1)}\right) \nonumber \\
&=\frac{\mo^2+\mf^2}{2\mo\mf}\Eon
+\frac{\mo^2-\mf^2}{2\mo\mf}\left(\frac{\mo}{2\rdt}-\dot{R}_{(1)}\right). \label{E4}
\end{align}
The crust of the gravastar is Shell 4 after the event $d_2$.  
As mentioned, since the gravastar becomes static after the event $d_2$, $\dot{R}_{(4)}$ always vanishes. 
Since we have
$$
\dot{R}_{(4)}^2=-U_{(4)}\left(\Rfo\right):=\left(\Efo+\frac{\mf}{2\Rfo}\right)^2-F_2\left(\Rfo\right), 
$$
Eq.~(\ref{E4}) and $\left. \dot{R}_{(4)}\right|_{\Rfo=\rdt}=0$ lead to 
\begin{equation}
\left[\frac{\mo^2+\mf^2}{2\mo\mf}\left(\Eon+\frac{\mo}{2\rdt}\right)
-\frac{\mo^2-\mf^2}{2\mo\mf}\dot{R}_{(1)}\right]^2-F_2=0. \label{static1}
\end{equation}
Since we have 
\begin{equation}
\dot{R}_{(1)}^2=\left(\Eon+\dfrac{\mo}{2\rdt}\right)^2-F_2=\left(\Eon-\dfrac{\mo}{2\rdt}\right)^2-F_1 \label{R1-2}
\end{equation}
at $\Ron=\rdt$, the following inequality holds; 
$$
\Eon+\frac{\mo}{2\rdt}>\left|\dot{R}_{(1)}\right|. 
$$
By the same argument as that of Eq.~(\ref{mass-order}), we have
\begin{equation}
\mo>\mf>0.   \label{mass-order2}
\end{equation}
Then, Eq.~(\ref{static1}) leads to
$$
\frac{\mo^2+\mf^2}{2\mo\mf}\left(\Eon+\frac{\mo}{2\rdt}\right)
-\frac{\mo^2-\mf^2}{2\mo\mf}\dot{R}_{(1)}-\sqrt{F_2}=0,
$$
and hence we have 
$$
\left(\Eon+\frac{\mo}{2\rdt}+\dot{R}_{(1)}\right)\mf^2-2\mo\mf\sqrt{F_2}+\left(\Eon+\frac{\mo}{2\rdt}-\dot{R}_{(1)}\right)\mo^2=0.
$$
The above quadratic equation for $\mf$ has a degenerate root
\begin{equation}
\mf=\frac{\mo\sqrt{F_2}}{\Eon+\dfrac{\mo}{2\rdt}+\dot{R}_{(1)}}, \label{m4}
\end{equation}
where we have used Eq.~(\ref{R1-2}).
By using Eq.~(\ref{R1-2}), we can see that Eq.~(\ref{mass-order2}) is satisfied only 
if $\dot{R}_{(1)}$ is positive. 
Thus, we consider the only situation in which $\dot{R}_{(1)}$ is positive at the event $d_2$. 
$\dot{R}_{(4)}$ vanishes if and only if the proper mass of Shell~4 satisfies Eq.~(\ref{m4}), 
and hereafter we assume so. By virtue of the future directed condition of the 
4-velosity of Shell~1, i.e., $\Eon-\dfrac{\mo}{2\rdt}>0$, and Eq.~(\ref{R1-2}), we have
$$
\Eon-\dfrac{\mo}{2\rdt}-\dot{R}_{(1)}>0. 
$$
Here note that Eq.~(\ref{E4}) can be rewritten as
$$
\Efo-\frac{\mf}{2\rdt}=\frac{\mo^2+\mf^2}{2\mo\mf}\left(\Eon-\frac{\mo}{2\rdt}-\dot{R}_{(1)}\right)
+\frac{\mo^2-\mf^2}{2\mf\rdt}+\frac{\mf}{\mo}\dot{R}_{(1)} . 
$$
Hence, if $\dot{R}_{(1)}>0$ holds, the future directed condition, $\Efo-\mf/2\rdt>0$, for Shell~4 also holds. 
The decay of Shell~1 to make the gravastar static is possible.

The effective potential of Shell~4, $U_{(4)}$, vanishes at $R_{(4)}=\rdt$ by assumption. 
The 1st and 2nd order derivatives of $U_{(4)}$ should vanish and be positive, respectively, 
at $R_{(4)}=\rdt$ so that the gravastar is stably static. Hereafter we assume so; these assumptions 
partly determine the equation of state of Shell~4 as follows.

Eq.~(\ref{thth-component}) implies that the surface energy density of Shell~4 is given in the form  
\begin{equation}
\sigma_{(4)}
=\frac{1}{4\pi\Rfo}\left[\sqrt{F_2\left(\Rfo\right)-\Ufo\left(\Rfo\right)}
-\sqrt{F_3\left(\Rfo\right)-\Ufo\left(\Rfo\right)}\right], \label{sigma4}
\end{equation}
and the tangential pressure of Shell~4, $P_{(4)}$, is given from Eq.~(\ref{E-con2}) in the form
\begin{align}
P_{(4)}&=-\frac{1}{2\Rfo}\frac{d}{d\Rfo}\left(\sigma_{(4)}\Rfo^2\right) \nonumber \\
&=-\left[\frac{1}{2}+\frac{\Rfo}
{4\sqrt{\left[F_2\left(\Rfo\right)-\Ufo\left(\Rfo\right)\right]\left[F_3\left(\Rfo\right)-\Ufo\left(\Rfo\right)\right]}}
\dfrac{d\Ufo\left(\Rfo\right)}{d\Rfo}\right]\sigma_{(4)} \nonumber \\
&+\frac{\Lambda_2\Rfo}{24\pi\sqrt{F_2\left(\Rfo\right)-\Ufo\left(\Rfo\right)}}
+\frac{M_3}{8\pi\Rfo^2\sqrt{F_3\left(\Rfo\right)-\Ufo\left(\Rfo\right)}}.
\label{P4}
\end{align}
At $\Rfo=\rdt$, we have
\begin{equation}
\left.\sigma_{(4)}\right|_{d_2}
=\frac{1}{4\pi\rdt}\left[\sqrt{F_2\left(\rdt\right)}-\sqrt{F_3\left(\rdt\right)}\right], \label{sigma-asym}
\end{equation}
and
\begin{align}
\left.P_{(4)}\right|_{d_2}&=-\frac{1}{2}\left.\sigma_{(4)}\right|_{d_2} 
+\frac{\Lambda_2\rdt}{24\pi\sqrt{F_2\left(\rdt\right)}}
+\frac{M_3}{8\pi\rdt^2\sqrt{F_3\left(\rdt\right)}}. \label{P-asym}
\end{align}
We are interested in whether the dominant energy condition 
$\sigma_{(4)}\geq\left|P_{(4)}\right|$ holds.

\subsubsection{The case of Shell~1 expanding at $d_1$}

First we consider the case of $\dot{R}_{(1)}>0$ at the first decay event $d_1$ with the areal radius $\rdo$. 
Since we consider the case that $\rdo$ is very close to $2M_1$, we have, from Eq.~(\ref{m1-exp}), 
\begin{equation}
\left.\mo\right|_{d_1}<\frac{\mz}{2\left(\Eze-\dfrac{1}{4\Eze}\right)}\left[1+{\cal O}\left(F_1\right)\right]\sqrt{F_1}. 
\label{m1-exp-2}
\end{equation}
The proper mass of Shell~1 should be much smaller than $\mz$. 

\begin{figure}[htbp]
 \begin{center}
 \includegraphics[width=12cm,clip]{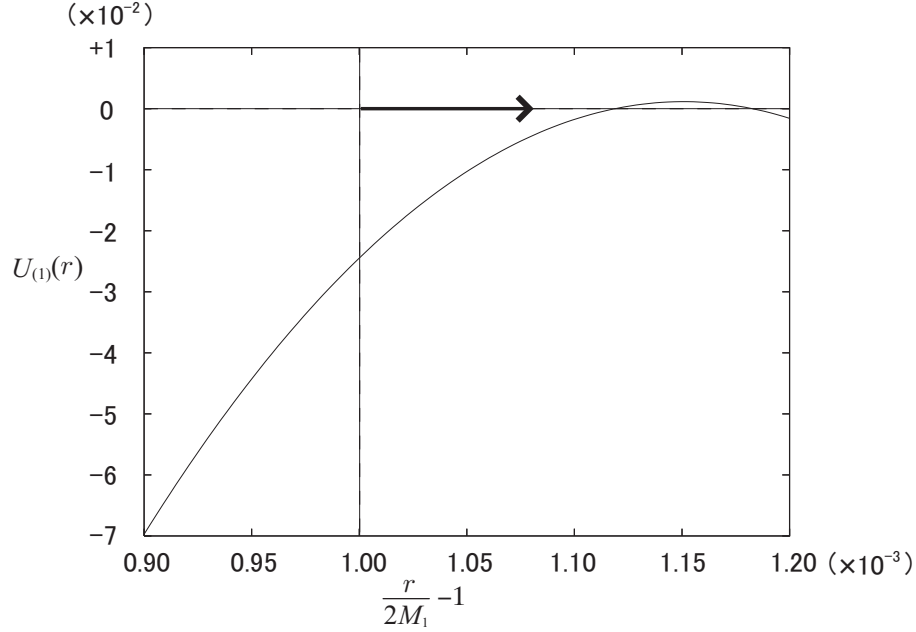}
 \end{center}
 \caption{
The effective potential $U_{(1)}$ near $\Ron=\rdo$ is depicted in the case that Shell~1 is the dust, i.e., 
$P_{(1)}=0$. We assume $\rdo=1.001 \times 2M_1$, $\Eze=0.9$ and $\mo=10^{-1}\mc$. 
Shell~1 begins expanding at $d_1$ and then bounces off the potential barrier. }
 \label{fig-potential-dust}
\end{figure}

We show the effective potential $\Uon$ in the case of the dust, $P_{(1)}=0$,  
in Fig.~\ref{fig-potential-dust}. Shell~1 will bounce off the 
potential barrier and then form a black hole by its contraction. 
The behavior of $\Uon$ even in the case 
\begin{equation}
P_{(1)}=w_{(1)}\sigma_{(1)} \label{EOS}
\end{equation}
with $\left|w_{(1)}\right| \leq 1$ 
is too similar to distinguish from that of the dust, even if it is depicted together in Fig.~\ref{fig-potential-dust}. 
The dominant energy condition for Shell~1 is given by 
\begin{equation}
\sigma_{(1)}\geq \left|P_{(1)}\right|. \label{DE-cond}
\end{equation}
As long as the dominant energy condition is satisfied, the effective potential of Shell~1 behaves 
as that shown in Fig.~\ref{fig-potential-dust}. 

As mentioned below Eq.~(\ref{R1-2}), 
Shell 1 should decay into Shell 3 and Shell 4, when $\dot{R}_{(1)}>0$ so that the gravastar is static. 
Hence Shell 1 should decay before it bounces off the potential barrier. The allowed domain 
for the motion of Shell 1 is bounded from above as Eq.~(\ref{Ru}) 
and hence $\rdo<\rdt<\Ru$. 

Let us estimate $\Ru$. 
Since $0<F_1(\rdo)\ll1$ and $\mo\ll\mz$ at $d_1$ due to Eq.~(\ref{m1-exp-2}), 
we have, from Eqs.~(\ref{calM2}) and (\ref{Lambda}), 
\begin{equation}
\Lambda_2=\frac{6M_1}{\rdo^3}\left[1-\beta^2-\frac{F_1}{4\Eze\left(\Eze-\dfrac{\mz}{2\rdo}\right)}
+{\cal O}\left(\beta^2 F_1,F_1^2\right)\right], \label{Lambda-2}
\end{equation}
where 
\begin{equation}
\beta:=\left.\frac{\mo}{\mz}\right|_{d_1} \leq{\cal O}\left(\sqrt{F_1}\right),   \label{beta}
\end{equation}
and hence $\Ru$ defined as Eq.~(\ref{Ru}) is written in the form,
\begin{equation}
\Ru=\rdo\left[1+\frac{\beta^2}{3}+\frac{F_1}{12\Eze\left(\Eze-\dfrac{\mz}{2\rdo}\right)}
+{\cal O}\left(\beta^4,\beta^2F_1,F_1^2\right)\right],
\label{Ru-2}
\end{equation}
where all quantities are evaluated at $d_1$. 
Since $\rdt<\Ru$ should be satisfied from Eq.~(\ref{Ru}), we have
\begin{equation}
\rdo<\rdt<\Ru=\rdo\left[1+{\cal O}\left(\beta^2,F_1\right)\right]. \label{order}
\end{equation} 
Since we consider the situation that $\rdo$ is larger than but very close 
to $2M_1$, $\rdt$ should be very close to $2M_1$ from 
Eq.~(\ref{order}). Furthermore, from Eq.~(\ref{Lambda-2}) 
and the inequality $0<\rdt-\rdo \leq {\cal O}\left(\beta^2,F_1\right)$ derived from Eq.~(\ref{order}), 
we have
$$
\frac{\Lambda_2}{3}\rdt^2=\frac{2\left.{\cal M}_{(2)}\right|_{d_1}}{\rdo^3}\rdt^2
=\frac{2M_1}{\rdt}+{\cal O}\left(\beta^2,F_1\right),
$$
and hence 
\begin{equation}
F_2\left(\rdt\right)={\cal O}\left(\beta^2,F_1\right)\ll1. \label{F2-d2}
\end{equation}
Here again note that $F_2\left(\rdt\right)>F_3\left(\rdt\right)$ holds because of $\Efo>0$: 
see Eq.~(\ref{E4}) and the discussion below Eq.~(\ref{m4}). 

From Eqs.~(\ref{sigma-asym}), (\ref{P-asym}) and (\ref{F2-d2}), 
Eqs.~(\ref{sigma-asym}) and (\ref{P-asym}) 
imply $\sigma_{(4)}\ll P_{(4)}$, and so  
the violation of the dominant energy condition (\ref{DE-cond}). 
This result is basically equivalent to that obtained by Visser and Wiltshire 
for their gravastar model\cite{VW2004}.

\subsubsection{The case of Shell~1 contracting at $d_1$}

\begin{figure}[htbp]
 \begin{center}
 \includegraphics[width=12cm,clip]{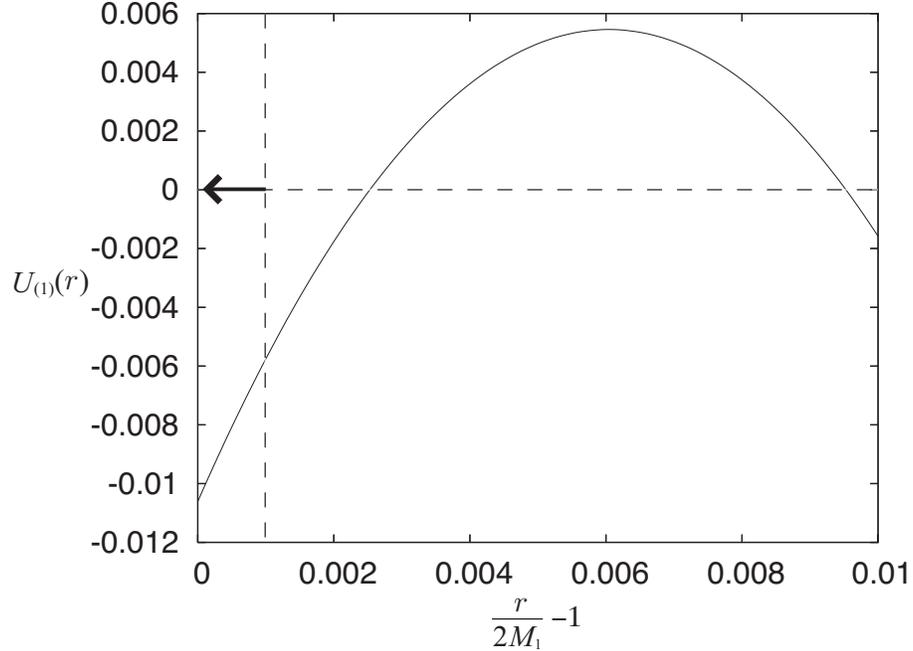}
 \end{center}
 \caption{
The same as Fig.~\ref{fig-potential-dust}, but $\mo=5\mc$. 
In this case, Shell~1 begins contracting at $d_1$ and then a black hole forms. }
 \label{fig-potential-dust-con}
\end{figure}

We consider the case that Shell~1 begins contracting at the first decay event $d_1$; the proper mass $\mo$ 
satisfies Eq.~(\ref{m1-con}). As in the expanding case, we show the effective potential $\Uon$ 
in the case of $P_{(1)}=0$ in Fig.~\ref{fig-potential-dust-con}. 
As in the case of expansion at $d_1$, $\Uon$ of the equation of state (\ref{EOS})  
with $\left|w_{(1)}\right|\leq1$ is too similar to that of the dust to distinguish between them, 
even if they are depicted together in Fig.~\ref{fig-potential-dust-con}. 
In this case, Shell~1 does not bounce off the 
potential barrier but directly forms a black hole through its contraction. 
Thus, in the contracting case, the equation of state of Shell~1 can not 
be Eq.~(\ref{EOS}) with $\left|w_{(1)}\right|\leq1$ so that the gravastar forms. 

\begin{figure}[htbp]
 \begin{center}
 \includegraphics[width=8cm,clip]{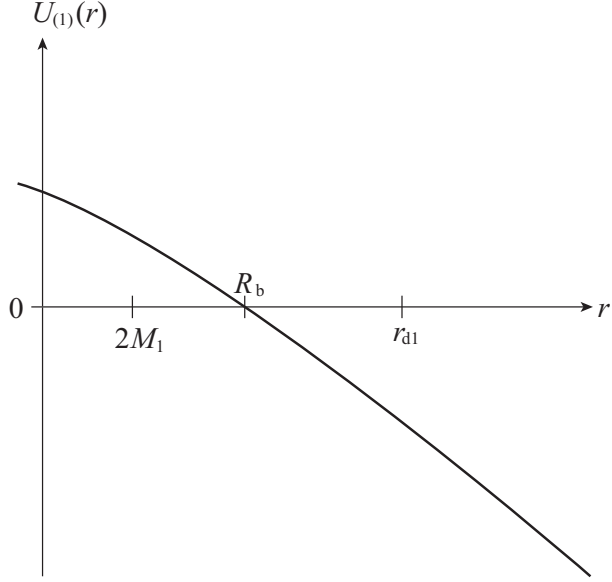}
 \end{center}
 \caption{
The assumed effective potential $U_{(1)}$ of Shell~1 near $R=R_{\rm b}$ is schematically 
depicted.  }
 \label{fig-potential}
\end{figure}

Shell~1 should bounce off the potential barrier at  some radius $\Rb$ larger than $2M_1$ 
so that the black hole formation is halted; the effective potential should take 
the following form near $d_1$;
\begin{equation}
\Uon(r)=-\alpha \left(r-\Rb\right)+{\cal O}\left(\left(r-\Rb\right)^{n}\right), \label{U1-2}
 \end{equation}
 where $\alpha$ and $n$ are positive constant and natural number larger than one, respectively, and  
 \begin{equation}
 2M_1<\Rb<\rdo \label{r-order}
 \end{equation}
 should hold (see Fig.~\ref{fig-potential}). 
 By contrast to the case of Shell~1 expanding at $d_1$, in the present case,  
 $\mo$ does not have to be much smaller than $\mz$ due to Eq.~(\ref{m1-con}) and we assume that 
 $\mo$ is close to but less than $\mz$.  
 Equation (\ref{calM2}) leads to 
\begin{align}
\left.{\cal M}_{(2)}\right|_{d_1}= \frac{\mz^2-\mo^2}{\mz^2}M_1+{\cal O}\left(F_1\right). 
\end{align}
Hence, Equation (\ref{Ru}) implies 
$$
\Ru=\frac{\rdo}{\left(1-\beta^2\right)^{1\over3}} \left[1+{\cal O}\left(F_1\right)\right],
$$ 
where $\beta$ has been defined as Eq.~(\ref{beta}) and is less than but can be very close to unity, 
and hence $\Ru$ may be much larger than $\rdo$. As a result, $\rdt$ can also be much larger than $\rdo$, 
and hence, as we will discuss later, the dominant energy condition (\ref{DE-cond}) can be satisfied by Shell 4. 
However,  as shown below, the dominant energy condition is not satisfied at $\Ron=\Rb$ by Shell~1. 

\begin{figure}[htbp]
 \begin{center}
 \includegraphics[width=12cm,clip]{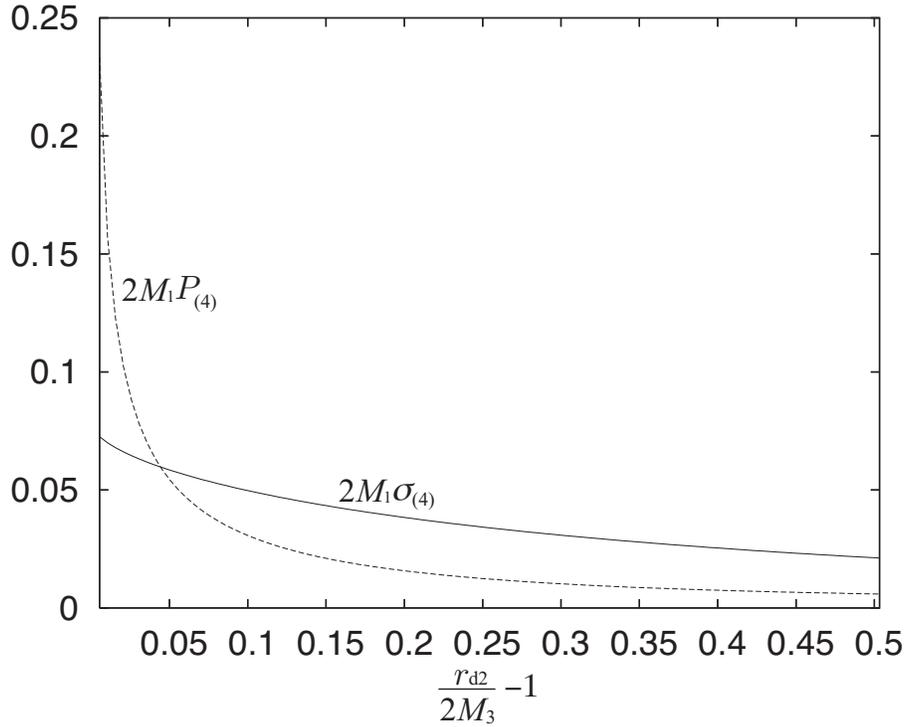}
 \end{center}
 \caption{
The surface energy density and tangential pressure of Shell~4  
are depicted as a function of the final radius of the gravastar $\Rfo=\rdt$. 
We assume $\Eze=0.9$, $\mo=0.99\mz$ at $d_1$ and $\rdo=1.00001\times 2M_1$, and 
$v=\sqrt{F_1(\rdo)}=3.16\times10^{-3}$. }
 \label{fig-DE-1}
\end{figure}

\begin{figure}[htbp]
 \begin{center}
 \includegraphics[width=10cm,clip]{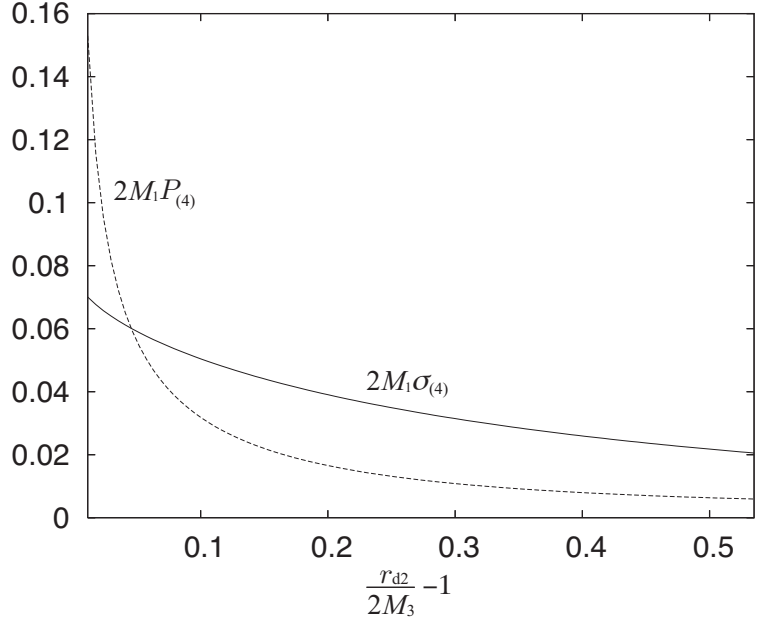}
 \end{center}
 \caption{
The same as Fig.~\ref{fig-DE-1} but $v=10\sqrt{F_1(\rdo)}=3.16\times10^{-2}$. }
 \label{fig-DE-10}
\end{figure}

\begin{figure}[htbp]
 \begin{center}
 \includegraphics[width=10cm,clip]{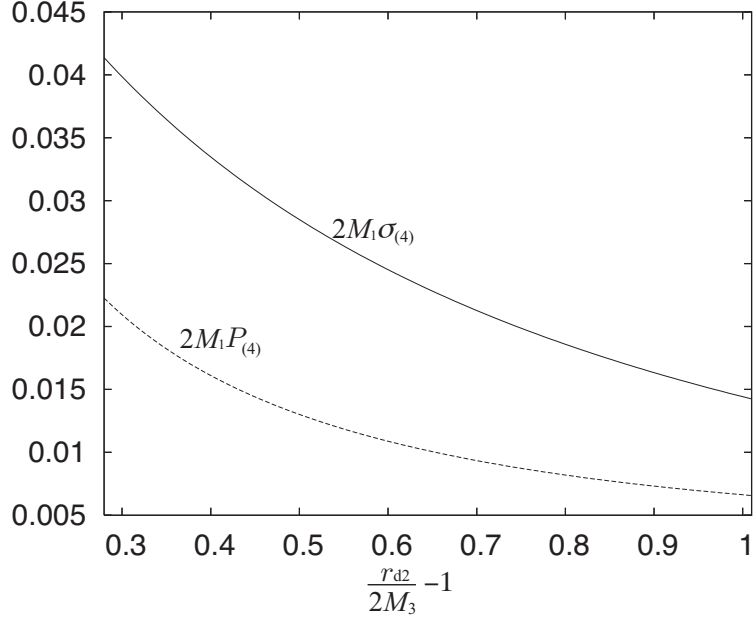}
 \end{center}
 \caption{
The same as Fig.~\ref{fig-DE-1} but $v=100\sqrt{F_1(\rdo)}=3.16\times10^{-1}$. From Fig.~\ref{fig-M3-r}, we 
can see $0.74M_1<M_3 < 0.78M_1$ in this case. Hence, the lower bound of the domain of this 
figure is restricted by a bit large value. }
 \label{fig-DE-100}
\end{figure}

\begin{figure}[htbp]
 \begin{center}
 \includegraphics[width=10cm,clip]{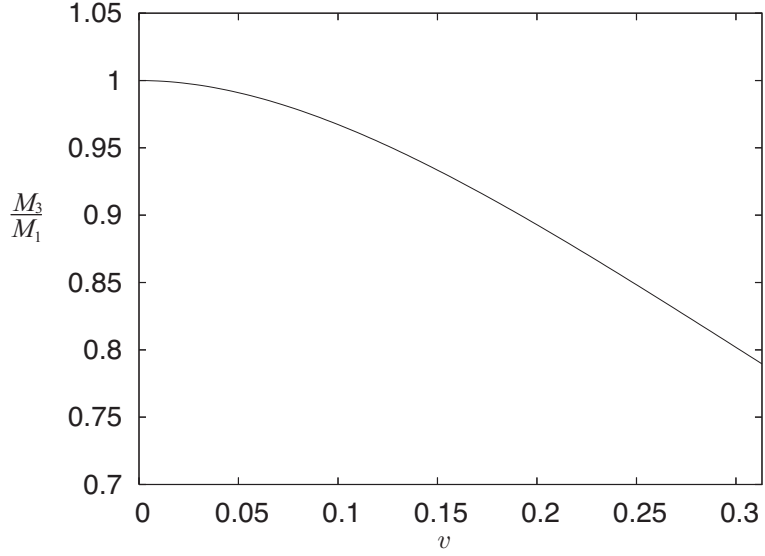}
 \end{center}
 \caption{
The mass parameter $M_3$ in the case of $\rdt=\rdo=1.00001\times 2M_1$ is depicted as a function of $v$. }
 \label{fig-M3-v}
\end{figure}

\begin{figure}[htbp]
 \begin{center}
 \includegraphics[width=10cm,clip]{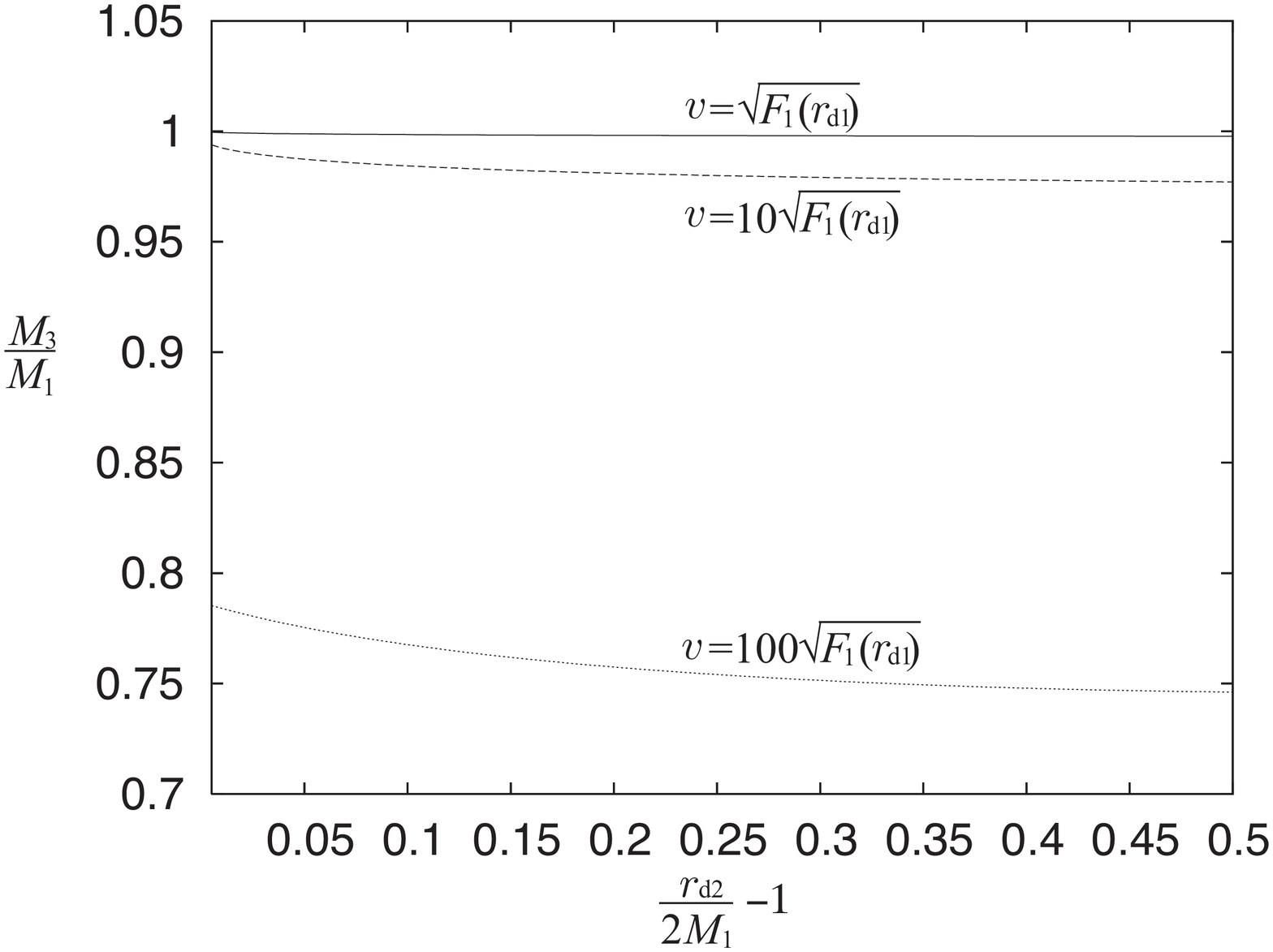}
 \end{center}
 \caption{
The mass parameter $M_3$ in the case of $v=\sqrt{F_1\left(\rdo\right)}$, $10\sqrt{F_1\left(\rdo\right)}$ and $100\sqrt{F_1\left(\rdo\right)}$ 
is depicted as a function of $\rdt$. 
As in Figs.~\ref{fig-DE-1}--\ref{fig-DE-100}, we assume $\Eze=0.9$, $\mo=0.99\mz$ at $d_1$ and $\rdo=1.00001\times 2M_1$.}. 
 \label{fig-M3-r}
\end{figure}

Through the same prescription to derive Eqs.~(\ref{sigma4}) and (\ref{P4}), we have 
\begin{equation}
\sigma_{(1)}=\frac{1}{4\pi\Ron}\left[\sqrt{F_2\left(\Ron\right)-\Uon\left(\Ron\right)}
-\sqrt{F_1\left(\Ron\right)-\Uon\left(\Ron\right)}\right], \label{sigma1}
\end{equation}
and
\begin{align}
P_{(1)}
&=-\frac{1}{2\Ron}\frac{d}{d\Ron}\left(\sigma_{(1)}\Ron^2\right) \nonumber \\
&=-\left[\frac{1}{2}+\frac{\Ron}
{4\sqrt{\left[F_2\left(\Ron\right)-\Uon\left(\Ron\right)\right]\left[F_1\left(\Ron\right)-\Uon\left(\Ron\right)\right]}}
\dfrac{d\Uon\left(\Ron\right)}{d\Ron}\right]\sigma_{(1)} \nonumber \\
&+\frac{\Lambda_2\Ron}{24\pi\sqrt{F_2\left(\Ron\right)-\Uon\left(\Ron\right)}}
+\frac{M_1}{8\pi\Ron^2\sqrt{F_1\left(\Ron\right)-\Uon\left(\Ron\right)}}. \label{P1}
\end{align}
Since we have
$$
\left.\frac{d\Uon\left(\Ron\right)}{d\Ron}\right|_{\Ron=\Rb} =-\alpha < 0,
$$
we obtain, from Eq.~(\ref{P1}), 
\begin{align}
\left.P_{(1)}\right|_{\Ron=\Rb}\geq-\frac{1}{2}\left.\sigma_{(1)}\right|_{\Ron=\Rb}
+\frac{\Lambda_2\Rb}{24\pi\sqrt{F_2\left(\Rb\right)}}
+\frac{M_1}{8\pi\Rb^2\sqrt{F_1\left(\Rb\right)}}. 
\end{align}
Due to Eq.~(\ref{r-order}) and since $\rdo$ is very close to $2M_1$, $F_1\left(\Rb\right)\ll1$ holds.  
Hence we have
\begin{align}
\left.P_{(1)}\right|_{\Ron=\Rb}
&\simeq \frac{M_1}{8\pi\Rb^2\sqrt{F_1\left(\Rb\right)}} \gg \left.\sigma_{(1)}\right|_{\Ron=\Rb}. \nonumber
\end{align}
The dominant energy condition can not be satisfied by Shell~1 at and around $\Ron=\Rb$ by continuity.


Now we see the equation of state of Shell~4 
which is the crust of the gravastar after the second decay event $d_2$; 
the surface energy density and the tangential pressure of Shell~4, 
$\sigma_{(4)}$ and $P_{(4)}$ are evaluated 
by using Eqs.~(\ref{sigma-asym}) and (\ref{P-asym}).  
Although we have determined the effective potential $\Uon$ of Shell~1 in the only vicinity of $\Ron=\Rb$ as Eq.~(\ref{U1-2}), 
we have not yet in the vicinity of $\Ron=\rdt$. Thus, the value of $\Uon$, or equivalently, $\dot{R}_{(1)}$ at $d_2$ 
is  regarded as a free parameter. Once we assume the values of $\rdt$ and 
$$
v:=\left.\dot{R}_{(1)}\right|_{d_2},
$$
we have
$$
\left.m_{(1)}\right|_{d_2}=\left.4\pi\sigma_{(1)}\Ron^2\right|_{d_2}
=\rdt\left[\sqrt{v^2+F_2(\rdt)}-\sqrt{v^2+F_1(\rdt)}\right]
$$
and
$$
\left.\Eon\right|_{d_2}=\frac{\rdt}{2\mo}\left[F_2(\rdt)-F_1\left(\rdt\right)\right].
$$

We depict $\sigma_{(4)}$ and $P_{(4)}$ as a function of $\rdt/2M_3-1$ 
in the case of $\Eze=0.9$, $\mo=0.99\mz$ at $d_1$ and $\rdo=1.00001\times 2M_1$ for three values of $v$, 
in Figs.~\ref{fig-DE-1}--\ref{fig-DE-100}.  

We also show $M_3$ in the case of $\rdt=\rdo=1.00001\time 2M_1$ as a function of $v$ in Fig.~\ref{fig-M3-v}; 
the larger $v$ is, the smaller $M_3$ is. This behavior implies that if Shell~1 has the larger outward velocity 
$v>0$, the larger energy should be released through the emission of Shell~3 so that Shell~4 is at rest. 
Furthermore, we depict $M_3$ as a function of $\rdt/2M_1-1$ in Fig.~\ref{fig-M3-r} for three values of $v$; 
here note that $\rdt$ is normalized by not $M_3$ but $M_1$.  The mass parameter $M_3$ is a 
decreasing function of $\rdt$. 

We can see from Figs.~\ref{fig-DE-1} and \ref{fig-DE-10} that  
the dominant energy condition $\sigma_{(4)}\geq\left|P_{(4)}\right|$ is satisfied 
in the case of $\rdt \gtrsim 1.04\times 2M_3$, 
it is not so for $\rdt$ very close to $2M_3$; the domain in Fig.~\ref{fig-DE-100} does not 
include  $\rdt = 1.04\times 2M_3$ due to the behavior of $M_3$ shown in Figs.~\ref{fig-M3-v} and \ref{fig-M3-r}.  
Since $\rdt$ is the radius of the gravastar in its final state, if $\rdt \gtrsim 1.04\times 2M_3$ holds, 
the formed gravastar satisfies the dominant energy condition, even though the crust of the gravastar 
does not in its formation process. 
The quantum gravitational effects should play an important role so that 
the process accompanied by the violation of the dominant energy condition is realized.  Hence 
the gravastar formation should rely on the quantum gravitational effects, 
if it begins at the very late stage of the gravitational collapse, i.e., $0<\rdo-2M_1\ll 2M_1$.

\begin{figure}[htbp]
 \begin{center}
 \includegraphics[width=6cm,clip]{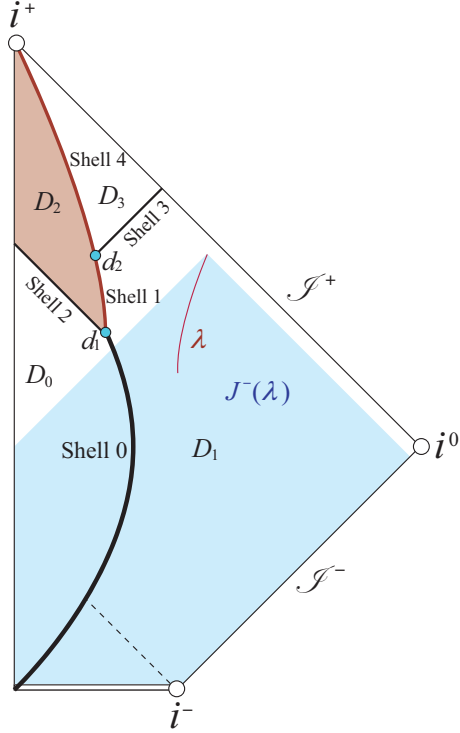}
 \end{center}
 \caption{
This is the case in which the observer $\lambda$ will wrongly 
conclude that a black hole will form  
if the areal radius $\rd$ at the beginning of the gravastar formation 
is sufficiently close to the gravitational radius $2M_1$ in $D_1$, 
but no event horizon forms. In the domain $D_2$ corresponds to the gravastar. 
Here Shell 0 is assumed to gravitationally contract, i.e.,  $1/2<\Eze<1$.   }
 \label{gravastar}
\end{figure}

As mentioned, it is observationally very important when the gravastar formation begins. 
If the gravastar formation starts after the backreaction of the Hawking radiation 
begins sufficiently affecting the evolution of the contracting object,  
the gravitationally contracting object of the mass larger than that of the solar mass 
will form a gravastar completely outside the causal past of observers 
with the finite lifetime like us. In such a case, the observers will wrongly 
conclude that a black hole will form (see Fig. \ref{gravastar}).

\section{Summary}

Any observer outside black holes cannot detect any physical signal caused by the 
black holes themselves but see the gravitationally contracting objects and phenomena 
caused by them; for  observers outside black holes, 
the contracting objects will form after infinite time lapses if they are the cases.  
In order to see why a contracting object seems to be a black hole even if 
there is not an event horizons but the contracting object in our view, 
we have studied a very simple model which describes the 
gravitational contraction of an infinitesimally thin spherical massive shell 
and studied null rays in such a situation. 
Even in the case that the shell made of materials which 
causes specular reflection of or is transparent to the null rays, 
it behaves as a black body due to its gravity if its radius is very close to its gravitational radius; 
incident null rays do not return from the shell or suffer indefinitely large redshift even if they return. 
Hence, the shell at the very late stage of its gravitational collapse is well approximated by the 
maximally extended Schwarzschild spacetime with the boundary condition 
under which nothing comes from the white hole: 
signals of the quasi-normal ringing and shadow images obtained in the spacetime with the 
shell will be, in practice, indistinguishable to those of the maximally extended Schwarzschild spacetime 
for any distant observer in the very late stage of the gravitational collapse. 
In this sense, the black hole shadow is not the appropriate name in the case of 
the observed black hole candidates, since it is not a shadow of a black hole 
but the image of the highly darkened photosphere of gravitationally contracting object. 
Even in the case of the black hole spacetime, the black hole shadow is not the appropriate name, 
since it is an image of the white hole.  

However, as we have shown, even though the observers detect the quasi-normal ringings 
and take photos of shadow images, those observables do not necessarily imply that 
the event horizon will form by the contracting object. There always remains the possibility 
that the formation of the event horizon is prevented by some unexpected event. 
We have given a scenario in which such a situation is realized: 
a gravitational contraction of the dust shell suddenly stops 
due to its decay into two daughter shells concentric 
with the parent shell, and then a gravastar forms. 
If the decay occurs at the radius so close to that of the corresponding event horizon that 
the decay event is outside of the causal past of observers, it may be impossible for the 
observers with finite lifetime to see the gravastar formation and hence such observers believe 
that the shell will form a black hole, even if there is no event horizon. 
On the other hand, our analysis on a simplified
formation scenario suggests that the formation of gravastar with the radius extremely close
to that of the would-be horizon may be possible only with large violation of the dominant
energy condition by the crust of the gravastar.  

\section{Some remarks}

Here we should note that the Hawking radiation can also not be the observable 
that is an evidence of the event horizon formation. 
As shown by  Paranjape and Padmanabhan, 
almost Planckian distribution of particles created through the semi-classical effect 
will appear in the contracting shell model~\cite{PP}, 
and hence the gravastar formation model cannot be distinguished from the black hole spacetime 
through the particle creation due to the semi-classical effects if the gravastar formation starts 
at the too late stage of the gravitational collapse to be observed by the distant observers 
with finite lifetimes. 
This issue will also be discussed by one of the present authors and his collaborators~\cite{HMC}.  
It might be interesting that the Planckian distribution is consistent to the approximate black body behavior of 
the shell at very late stage of its gravitational collapse. 

As mentioned, the gravastar formation might start after the effect of the Hawking radiation causes 
significant backreaction effects on the gravitational collapse of a massive object. If it is really so, 
the gravitational collapse of the massive object with the mass larger than the solar mass 
will not cause the gravastar formation within the age of the universe. 
By contrast, the formation of the primordial black hole with the mass much smaller than the solar mass 
should be replaced by the primordial gravastar formation that is, in principle, observable for us\cite{Chirenti-Rezzolla2007,Chirenti-Rezzolla2016}. 
Although it is very difficult to observe compact objects with very small mass, they might be very important 
in order to find the unexpected events. 

Rigorously speaking, it is impossible for us to conclude, through any observation, that it is a black hole.   
It is a profound fact that general relativity has predicted the advent of domains of 
which the existence can not be confirmed through any observation.  
By contrast, if it is not a black hole, we can, in principle, know that it is the case. 
It is necessary to keep observing black hole candidates.

\section*{Acknowledgments}
K.N. is grateful to Hideki Ishihara, Hirotaka Yoshino and colleagues at the elementary particle physics 
and gravity group in Osaka City University. T.H. is grateful to V. Cardoso for fruitful discussion on gravastars. 
This work was supported by JSPS KAKENHI Grant Number JP16K17688 (C.Y.).

\appendix

\section{Redshift of null ray due to massive spherical shell}

Here we consider the redshift of a null ray due to a spherical massive shell considered in Sec.~II; notation
adopted in this section is the same as that in Sec.~II. The null ray goes along a null geodesic. 
The components of the null geodesic tangent $l^a$ are written in the form
\begin{equation}
l^{\mu}_\pm=\omega_\pm\left(\frac{1}{F_\pm(r)},~\epsilon \sqrt{1-\frac{b_\pm^2}{r^2}F_\pm(r)},
~0,~\frac{b_\pm}{r^2}\right),  \label{null}
\end{equation}
where $\omega_\pm$ and $b_\pm$ are constants corresponding to the angular frequency and 
the impact parameter, respectively, and $\epsilon=\pm1$: $\epsilon=+1$ for the outgoing null, 
whereas $\epsilon=-1$ for the ingoing one. Without loss of generality, we consider the only case of 
the non-negative impact parameter $b_\pm\geq0$. 

In this section, for simplicity, we focus on the shell with no electric charge in the spacetime without the 
cosmological constant and the domain $D_-$ is Minkowskian;
$$
F_+=1-\frac{2M_+}{r}~~~~{\rm and}~~~~F_-=1.
$$
We also focus on the case that the spherical massive shell is contracting $\dot{R}<0$. 
 
\begin{figure}[htbp]
 \begin{center}
 \includegraphics[width=9cm,clip]{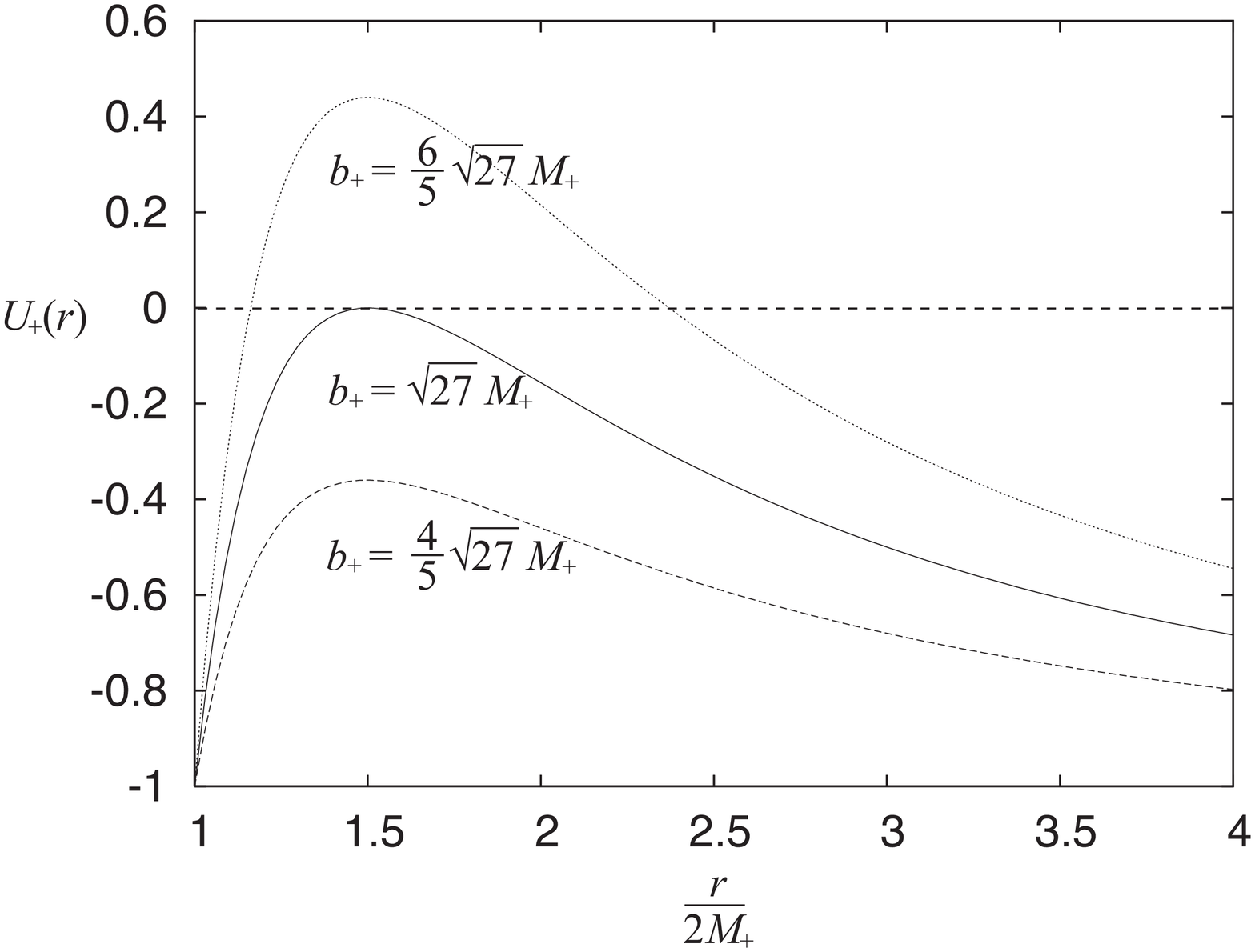}
 \end{center}
 \caption{
The effective potential of the null ray in the domain $D_+$ whose geometry is Schwarzschildian 
is depicted for three cases, $b_+=\dfrac{4}{5}\sqrt{27}M_+$, $b_+=\sqrt{27}M_+$ 
and $b_+=\dfrac{6}{5}\sqrt{27}M_+$. 
 }
 \label{null-potential}
\end{figure}

We obtain the energy equation from the radial component of $l_\pm^a$ as
\begin{equation}
\frac{1}{\omega_\pm^2}\left(\frac{dr}{d\gamma_\pm}\right)^2+W_\pm(r)=0,
\end{equation}
where $\gamma_\pm$ is the affine parameter and
\begin{equation}
W_\pm(r)=\frac{b_\pm^2}{r^2}F_\pm(r)-1.
\end{equation}
The null ray can move only in the domain of $W_\pm(r)\leq0$. 
The geometry of $D_+$ is Schwarzschildian, and as well known, the effective potential 
$W_+(r)$ has one maximum at $r=3M_+$ (see Fig.~\ref{null-potential}). 
If $b_+$ is larger than $\sqrt{27}M_+$, 
the maximum of $W_+$ is positive; the null ray going inward in the region of $r>3M_+$ 
bounces off the potential barrier and then goes away to infinity, whereas one going outward in 
the domain of $r<3M_+$ also bounces off the potential barrier and then turns to the center. 
On the other hand, the maximum of $W_+(r)$ is non-positive in the case of $b_+\leq\sqrt{27}M_+$;   
in this case, the null ray does not bounce off the potential barrier. The fact we should remember here 
is that if the null ray is ingoing, or equivalently, $\epsilon=-1$, 
in the region of $r<3M_+$ within $D_+$, 
it does not bounce off but continues to go inward. 

\subsection{Reflected case}

Let us consider the case that an ingoing null ray $l_a^{\rm (in)}$ from infinity in $D_+$  
with $\omega_+=\omi$ and $b_+=b_{\rm i}$ is reflected at the shell 
and then goes away to infinity in $D_+$ as an outgoing null ray  
$l_a^{\rm (out)}$ with $\omega_+=\omo$ and $b_+=b_{\rm o}$. 
Since the angular frequency of the reflected null ray 
is the same as that before reflection in the rest frame of the shell, we have
\begin{equation}
l_a^{\rm (in)}u^a=l _a^{\rm (out)}u^a. \label{reflection}
\end{equation}
The parameter $\epsilon$ of ingoing null ray should be equal to $-1$, 
whereas it is non-trivial which sign of $\epsilon$ is chosen after the reflection; 
$\epsilon$ after the reflection is denoted by $\epsilon_{\rm o}$. 
Then Eq.~(\ref{reflection}) leads to
\begin{align}
&\omi\left(\sqrt{1+\frac{F_+}{V^2}}
-\sqrt{1-\frac{\bi^2}{R^2}F_+} \right)
=\omo\left(\sqrt{1+\frac{F_+}{V^2}}
+\epsilon_{\rm o}\sqrt{1-\frac{\bo^2}{R^2}F_+}\right), \label{reflection2}
\end{align}
where $F_+=F_+(R)$, and
\begin{equation}
V:=-\dot{R}>0. \label{V-def}
\end{equation}


In the rest frame, the component of $l^a$ vertical to the shell changes its sign at the reflection event; 
\begin{equation}
l_a^{\rm (in)}n^a=-l _a^{\rm (out)}n^a. \label{reflection3}
\end{equation}
Equation (\ref{reflection3}) leads to
\begin{align}
&\omi\left[1-\sqrt{\left(1+\frac{F_+}{V^2}\right)
\left(1-\frac{\bi^2}{R^2}F_+\right)} \right]
=-\omo\left[1+\epsilon_{\rm o}\sqrt{\left(1+\frac{F_+}{V^2}\right)
\left(1-\frac{\bo^2}{R^2}F_+\right)} \right]. \label{reflection4}
\end{align}
On the other hand, the component of $l^a$ tangential to the shell does not change; the equation
$$
l_a^{\rm (in)}\hat{\phi}^a=l _a^{\rm (out)}\hat{\phi}^a
$$
leads to the conservation of the angular momentum, 
$$
\omi\bi=\omo\bo=:L.
$$

Since the null ray is assumed to hit the shell, $l_a^{\rm (in)}n^a<0$ should be satisfied, and hence  
\begin{equation}
\bi^2<\frac{R^2}{V^2+F_+} \label{hitting}
\end{equation}
should hold. 


From Eqs.~(\ref{reflection2}) and (\ref{reflection4}), we have
\begin{equation}
\frac{\sqrt{1+\dfrac{F_+}{V^2}}-\sqrt{1-\dfrac{\bi^2}{R^2}F_+}}
{1-\sqrt{\left(1+\dfrac{F_+}{V^2}\right)\left(1-\dfrac{\bi^2}{R^2}F_+\right)}}
=-\frac{\sqrt{1+\dfrac{F_+}{V^2}}+\epsilon_{\rm o}\sqrt{1-\dfrac{\bo^2}{R^2}F_+}}
{1+\epsilon_{\rm o}\sqrt{\left(1+\dfrac{F_+}{V^2}\right)\left(1-\dfrac{\bo^2}{R^2}F_+\right)}}.
\label{impact}
\end{equation}
We rewrite Eq.~(\ref{impact}) in the form, 
\begin{align}
A_{\rm L}\sqrt{1-\dfrac{\bo^2}{R^2}F_+}=\epsilon_{\rm o}A_{\rm R}, \label{impact2}
\end{align}
where
\begin{align}
A_{\rm L}&=1+\frac{F_+}{2V^2}-\sqrt{\left(1+\dfrac{F_+}{V^2}\right)\left(1-\dfrac{\bi^2}{R^2}F_+\right)}, \\
A_{\rm R}&=\left(1+\frac{F_+}{2V^2}\right)\sqrt{1-\dfrac{\bi^2}{R^2}F_+}-\sqrt{1+\dfrac{F_+}{V^2}}.
\end{align}
It is not so difficult to see that $A_{\rm L}>0$ whereas the sign of $A_{\rm R}$ depends on 
$\bi$, $R$ and $V$. Since the l.h.s. of Eq.~(\ref{impact2}) is positive, $\epsilon_{\rm o}$ should be 
chosen so that the r.h.s. is also positive, and hence we have 
\begin{equation}
\epsilon_{\rm o}=
\left\{
\begin{array}{ll}
+1&~~{\rm for}~~\bi^2<b_{\rm cr}^2,  \\
-1&~~{\rm for}~~\bi^2>b_{\rm cr}^2, \\
\end{array}\right.
\label{epsilon-o}
\end{equation}
where
$$
b_{\rm cr}^2:=\dfrac{R^2F_+}{\left(2V^2+F_+\right)^2}.
$$
The null ray with $\epsilon_{\rm o}=-1$ goes inward although it is the null ray reflected by the shell.  
Since we consider the case that the reflection occurs when the radius 
of the shell is very close to the gravitational radius $2M_+$, 
the reflected null ray with $\epsilon_{\rm o}=-1$ continues to move inward in $D_+$; 
in other words, the distant observers recognize the shell as an absorber of all null rays hitting the shell. 

By taking the square of Eq.~(\ref{impact2}) and using the relation
$$
 A_{\rm L}^2 - A_{\rm R}^2= \frac{\bi^2F_+^3}{4 V^4 R^2}, 
$$
we obtain
$$
\omo=\frac{2\omi}{F_+}V^2A_{\rm L}.
$$
Then, by regarding $\omo$ as a function of $L$, $\omi$, $R$ and $V$, we have
$$
\frac{\partial \omo}{\partial L}= \frac{2\omi}{F_+}V^2\frac{\partial A_{\rm L}}{\partial L}
=\frac{2\bi V^2}{R^2}\sqrt{\frac{1+\dfrac{F_+}{V^2}}{1-\dfrac{\bi^2}{R^2}F_+}}>0.
$$
This result implies that $\omo$ of the reflected null ray that can go away is bounded from the value 
with $L=\omi b_{\rm cr}$. 
When $\bi$ is equal to $b_{\rm cr}$,  $A_{\rm R}$ vanishes, and hence  
$$
1-\dfrac{\bo^2}{R^2}F_+=0
$$
holds; the reflected null ray has vanishing radial component of $l^a$. 
In this case, Eq.~(\ref{reflection2}) leads to 
$$
\omo|_{\bi=b_{\rm cr}}=\frac{\omi F_+}{2V^2+F_+}.
$$
Hence, the angular frequency of the reflected null ray is bounded from the above as
$$
\omo<\frac{\omi F_+}{2V^2+F_+}.
$$
For $0<F_+\ll1$, the reflected null ray with $\epsilon_{\rm o}=+1$ 
suffers indefinitely large redshift, i.e., $\omo\ll\omi$. 
Note that, in the case of $V=0$, i.e., the static shell, $\omo=\omi$. The redshift of the reflected null ray 
is caused by the contraction of the shell.

\subsection{Transmitted case}

We study the redshift of a null ray in the case that it is transmitted through the shell. 
The null ray is assumed to be in $D_+$ initially, enter $D_-$, and then 
return to $D_+$. We are interested in the case that when the null ray returns from $D_-$ to $D_+$, 
the radius $R$ of the shell is very close to the gravitational radius $2M_+$; here our attention 
is concentrated at the moment of this return. The angular frequency and the impact parameter 
of the null ray in $D_-$ are denoted by $\omm$ and $\bmi$, respectively, whereas  
those of the null ray after returning to $D_+$ are denoted by $\omp$ and $\bpu$, respectively. 

The inequality $l_an^a>0$, or equivalently,  
\begin{equation}
1+\epsilon\sqrt{\left(1+\frac{1}{V^2}\right)\left(1-\frac{\bmi^2}{R^2}\right)}>0 \label{hitting2}
\end{equation}
should hold just before the null ray hits the shell in $D_-$, where $V$ has been defined as 
Eq.~(\ref{V-def}). 
Equation~(\ref{hitting2}) is necessarily satisfied if $\epsilon$ is equal to unity. On the other hand, 
\begin{equation}
\bmi>\frac{R}{\sqrt{V^2+1}}
\end{equation}
should hold in the case of $\epsilon=-1$. We will study these cases separately. 
But in both cases, the continuity of $l_a\hat{\phi}^a$ leads to
$$
\omm \bmi=\omp \bpu.
$$
Since it is non-trivial whether $\epsilon$ is equal to $+1$ after entering $D_+$, 
$\epsilon$ in $D_+$ is denoted by $\epsilon_+$.

First, we consider the case of $\epsilon=-1$ in $D_-$. 
In the transmitted case, all components of $l^a$ should be everywhere continuous; 
the continuity of $l_a u^a$ leads to
\begin{equation}
\omm\left[\sqrt{1+\frac{1}{V^2}}-\sqrt{1-\dfrac{\bmi^2}{R^2}}\right]
=\frac{\omp}{F_+}\left[\sqrt{1+\dfrac{F_+}{V^2}}
+\epsilon_+\sqrt{1-\frac{\bpu^2}{R^2}F_+}\right], \label{trans-o1-m}
\end{equation}
whereas the continuity of $l_an^a$ leads to
\begin{equation}
\omm \left[1-\sqrt{\left(1+\frac{1}{V^2}\right)\left(1-\dfrac{\bmi^2}{R^2}\right)}\right]
=\frac{\omp}{F_+}\left[1+\epsilon_+\sqrt{\left(1+\dfrac{F_+}{V^2}\right)
\left(1-\frac{\bpu^2}{R^2}F_+\right)}\right], \label{trans-o2-m}
\end{equation}
where $F_+=F_+(R)$. 
As in the reflected case, by dividing each side of Eq.~(\ref{trans-o1-m}) 
by each side of Eq.~(\ref{trans-o2-m}) and further by a few manipulations, we have
\begin{equation}
B_{\rm L}\sqrt{1-\frac{\bpu^2}{R}F_+}=\epsilon_+B_{\rm R}, \label{trans-o3-m}
\end{equation}
where
\begin{align}
B_{\rm L}&=\sqrt{1+\frac{1}{V^2}}
\left(\sqrt{1+\frac{F_+}{V^2}}+\sqrt{1-\frac{\bmi^2}{R^2}}\right)
-\sqrt{\left(1+\frac{F_+}{V^2}\right)\left(1-\frac{\bmi^2}{R^2}\right)}-1, \\
B_{\rm R}&=\sqrt{1+\frac{F_+}{V^2}}+\sqrt{1-\frac{\bmi^2}{R^2}}
-\sqrt{1+\frac{1}{V^2}}
\left[\sqrt{\left(1+\frac{F_+}{V^2}\right)\left(1-\frac{\bmi^2}{R^2}\right)}+1\right]. \label{CR-m}
\end{align}
Since we have
\begin{align}
&\left(1+\frac{1}{V^2}\right)\left(\sqrt{1+\frac{F_+}{V^2}}+\sqrt{1-\frac{\bmi^2}{R^2}}\right)^2
-\left[\sqrt{\left(1+\frac{F_+}{V^2}\right)\left(1-\frac{\bmi^2}{R^2}\right)}+1\right]^2 \nonumber \\
=&\frac{1}{V^2}\left(\sqrt{1+\frac{F_+}{V^2}}+\sqrt{1-\frac{\bmi^2}{R^2}}\right)^2
+\frac{\bmi^2F_+}{V^2R^2}>0,
\end{align}
$B_{\rm L}>0$ holds. We also have
\begin{align}
&\left(\sqrt{1+\frac{F_+}{V^2}}+\sqrt{1-\frac{\bmi^2}{R^2}}\right)^2
-\left(1+\frac{1}{V^2}\right)
\left[\sqrt{\left(1+\frac{F_+}{V^2}\right)\left(1-\frac{\bmi^2}{R^2}\right)}+1\right]^2 \nonumber \\
=&
-\frac{1}{V^2}\left[\sqrt{\left(1+\frac{F_+}{V^2}\right)\left(1-\frac{\bmi^2}{R^2}\right)}+1\right]^2
+\frac{\bmi^2F_+}{V^2R^2}
\nonumber \\
=&-\frac{1}{V^2}
\left[\sqrt{\left(1+\frac{F_+}{V^2}\right)\left(1-\frac{\bmi^2}{R^2}\right)}+1+\frac{\bmi\sqrt{F_+}}{R}\right] 
\left[\sqrt{\left(1+\frac{F_+}{V^2}\right)\left(1-\frac{\bmi^2}{R^2}\right)}+1
-\frac{\bmi\sqrt{F_+}}{R}\right] \nonumber \\
=&-\frac{1}{V^2}
\left[\sqrt{\left(1+\frac{F_+}{V^2}\right)\left(1-\frac{\bmi^2}{R^2}\right)}+1
+\frac{\bmi\sqrt{F_+}}{R}\right] \nonumber \\
&\times\left[\sqrt{\left(1+\frac{F_+}{V^2}\right)\left(1-\frac{\bmi^2}{R^2}\right)}+1
-\frac{\bmi}{R}+\frac{\bmi}{R}\left(1-\sqrt{F_+}\right)\right]<0,
\end{align}
where, in the last inequality, we have used the fact that $\bmi/R\leq1$ and $\sqrt{F_+}<1$, and  
hence $B_{\rm R}<0$ holds. Through Eq.~(\ref{trans-o1-m}), this result implies 
that $\epsilon_+$ is equal to $-1$. This result implies that in the case of $\epsilon_-=-1$, 
the null ray keeps going inward even after returning to $D_+$ and is effectively absorbed 
by the contracting shell.

Next, we consider the case of $\epsilon=+1$ in $D_-$. 
By the similar argument to the case of $\epsilon=-1$ in $D_-$, 
the continuity of $l_a u^a$ leads to
\begin{equation}
\omm\left[\sqrt{1+\frac{1}{V^2}}+\sqrt{1-\dfrac{\bmi^2}{R^2}}\right]
=\frac{\omp}{F_+}\left[\sqrt{1+\dfrac{F_+}{V^2}}
+\epsilon_+\sqrt{1-\frac{\bpu^2}{R^2}F_+}\right], \label{trans-o1}
\end{equation}
whereas the continuity of $l_an^a$ leads to
\begin{equation}
\omm \left[1+\sqrt{\left(1+\frac{1}{V^2}\right)\left(1-\dfrac{\bmi^2}{R^2}\right)}\right]
=\frac{\omp}{F_+}\left[1+\epsilon_+\sqrt{\left(1+\dfrac{F_+}{V^2}\right)
\left(1-\frac{\bpu^2}{R^2}F_+\right)}\right].  \label{trans-o2}
\end{equation}
By dividing each side of Eq.~(\ref{trans-o1}) by each side of Eq.~(\ref{trans-o2}) and by 
a few simple manipulations, we have
\begin{equation}
C_{\rm L}\sqrt{1-\frac{\bpu^2}{R^2}F_+}=\epsilon_+C_{\rm R}, \label{trans-o3}
\end{equation}
where
\begin{align}
C_{\rm L}&=\sqrt{1+\frac{F_+}{V^2}}
\left(\sqrt{1+\frac{1}{V^2}}+\sqrt{1-\frac{\bmi^2}{R^2}}\right)
-\sqrt{\left(1+\frac{1}{V^2}\right)\left(1-\frac{\bmi^2}{R^2}\right)}-1, \\
C_{\rm R}&=-\sqrt{1+\frac{1}{V^2}}-\sqrt{1-\frac{\bmi^2}{R^2}}
+\sqrt{1+\frac{F_+}{V^2}}
\left[\sqrt{\left(1+\frac{1}{V^2}\right)\left(1-\frac{\bmi^2}{R^2}\right)}+1\right]. \label{CR}
\end{align}
By the similar prescription to that in the case of $B_{\rm L}$, we can see $C_{\rm L}>0$.  
Hence, $C_{\rm R}$ should be positive so that $\epsilon_+$ is equal to $+1$,  
although the sign of $C_{\rm R}$ is non-trivial. In order to know it, we study the 
following quantity
\begin{align}
{\cal C}_{\rm R}&:=
\left(1+\frac{F_+}{V^2}\right)
\left[\sqrt{\left(1+\frac{1}{V^2}\right)\left(1-\frac{\bmi^2}{R^2}\right)}+1\right]^2
-\left(\sqrt{1+\frac{1}{V^2}}+\sqrt{1-\frac{\bmi^2}{R^2}}\right)^2 \nonumber \\
&=\frac{F_+}{V^2}
\left[\sqrt{\left(1+\frac{1}{V^2}\right)\left(1-\frac{\bmi^2}{R^2}\right)}+1\right]^2
-\frac{\bmi^2}{V^2R^2} \nonumber \\
&=\frac{F_+}{V^2}\left[\sqrt{\left(1+\frac{1}{V^2}\right)\left(1-\frac{\bmi^2}{R^2}\right)}+1
+\frac{\bmi}{R\sqrt{F_+}}\right] \nonumber \\
&\times\left[\sqrt{\left(1+\frac{1}{V^2}\right)\left(1-\frac{\bmi^2}{R^2}\right)}+1
-\frac{\bmi}{R\sqrt{F_+}}\right].
\end{align}
$C_{\rm R}$ is positive, if and only if ${\cal C}_{\rm R}$ is positive. 
We can see that ${\cal C}_{\rm R}$ is positive if and only if 
\begin{equation}
\sqrt{\left(1+\frac{1}{V^2}\right)\left(1-\frac{\bmi^2}{R^2}\right)}+1
-\frac{\bmi}{R\sqrt{F_+}}>0 \label{escape}
\end{equation}
holds. If 
\begin{equation}
\frac{\bmi}{R\sqrt{F_+}}\leq1 \label{escape2}
\end{equation}
is satisfied, Eq.~(\ref{escape}) holds. By contrast, in the case of 
\begin{equation}
\frac{\bmi}{R\sqrt{F_+}}>1, \label{escape3}
\end{equation}
we rewrite Eq.~(\ref{escape}) in the form
$$
\sqrt{\left(1+\frac{1}{V^2}\right)\left(1-\frac{\bmi^2}{R^2}\right)}>\frac{\bmi}{R\sqrt{F_+}}-1
$$
and take the square of its both sides and, as a result, obtain 
$$
\left[\left(V^2+1\right)F_+ +V^2\right]\frac{\bmi^2}{R^2}-2V^2\sqrt{F_+}\frac{\bmi}{R}-F_+<0.
$$
Then, we have
\begin{equation}
\frac{V^2\sqrt{F_+}-\sqrt{\left(V^2+1\right)\left(V^2+F_+\right)F_+}}
{\left(V^2+1\right)F_++V^2}
<\frac{\bmi}{R}
<\frac{V^2\sqrt{F_+}+\sqrt{\left(V^2+1\right)\left(V^2+F_+\right)F_+}}
{\left(V^2+1\right)F_++V^2} \label{escape4}
\end{equation}
In order that the intersection between Eqs.~(\ref{escape3}) and (\ref{escape4}) is not empty, 
$$
\sqrt{F_+}<\frac{V^2\sqrt{F_+}+\sqrt{\left(V^2+1\right)\left(V^2+F_+\right)F_+}}
{\left(V^2+1\right)F_+ +V^2}
$$
should hold. We are interested in the case of $F_+\ll1$ in which this inequality 
is satisfied. Hence, in the case of $F_+\ll1$, $C_{\rm R}$ is 
positive if and only if
\begin{equation}
\bmi< b_{\rm cr}
\end{equation}
holds, where
\begin{align}
b_{\rm cr}&:=\frac{R\left[V^2\sqrt{F_+}
+\sqrt{\left(V^2+1\right)\left(V^2+F_+\right)F_+}\right]}
{\left(V^2+1\right)F_++V^2} 
\end{align}
holds. 

By using
\begin{equation}
C_{\rm L}^2 - C_{\rm R}^2 = \frac{b_-^2F_+}{V^4 R^2},  \label{CL2-CR2}
\end{equation}
we have from Eq.~(\ref{trans-o3}) 
\begin{equation}
\omp=\omm V^2 C_{\rm L}.  \label{omp-sol}
\end{equation}
As in the reflected case, by regarding $\omp$ as a function of $L$, $\omm$, $R$ and $V$, 
we have
$$
\frac{\partial \omp}{\partial L}=\omm V^2 \frac{\partial C_{\rm L}}{\partial L}
=\frac{\bmi V^2}{R^2}\frac{\sqrt{1+\dfrac{1}{V^2}}-\sqrt{1+\dfrac{F_+}{V^2}}}
{\sqrt{{1-\dfrac{\bmi^2}{R^2}}}}>0.
$$ 
Hence, $\omp$ of the null ray with $\epsilon_+=+1$ that can escape to infinity is 
bounded from above by the value of $\omp|_{\bmi=b_{\rm cr}}$. 
Since $C_{\rm R}|_{\bmi=b_{\rm cr}}=0$ holds, we have, from Eq.~(\ref{CL2-CR2}), 
\begin{equation}
C_{\rm L}|_{\bmi=b_{\rm cr}}=\frac{b_{\rm cr}\sqrt{F_+}}{V^2 R}. \label{CL-crit}
\end{equation}
Substituting Eq.~(\ref{CL-crit}) into Eq.~(\ref{omp-sol}), we have
$$
\omp|_{\bmi=b_{\rm cr}}
=\omm \frac{b_{\rm cr}\sqrt{F_+}}{R},
$$
and hence, in the case of $V>0$, 
$$
\omp<\omp|_{\bmi=b_{\rm cr}}
=\frac{\omm F_+}{V\left(\sqrt{V^2+1}-V\right)}\left[1+{\cal O}\left(F_+\right)\right].
$$
This result implies that the transmitted null ray going away to infinity suffers indefinitely large 
redshift in the limit of $F_+\rightarrow0$.   
Note that in the case of $V=0$, i.e., the static shell, the angular frequency $\omega_+$ 
of the transmitted null ray is the same as that of the incident null ray, in $D_+$. 
The redshift of the transmitted null ray is caused by the contraction of the shell.

\section{The conservation of the four-momentum}

We show that the ``four-momentum conservation" (\ref{MC-2})  is consistent with the Bianchi identity 
$\nabla_aT^{ab}=0$. The stress-energy tensor of Shell~$I$ ($I=0,2,3$) is written in the form
$$
T_{(I)}^{ab}=S_{(I)}^{ab}\delta\left(\chi_{(I)}\right),
$$
where $\delta(x)$ is Dirac's delta function, and $\chi_{(I)} $ is the Gaussian normal 
coordinate: Shell~$I$ is located at $\chi_{(I)}=0$.

We introduce a coordinate system $(\tau,\chi, \hat{\theta},\hat{\phi})$ for the 
neighborhood of the decay event $d$ to which the coordinates $(\tau,\chi)=(0,0)$ is assigned.  
The coordinate $\chi$ is the Gaussian normal coordinate associated to the hypersurface  $\varSigma$ 
that agrees with the world hypersurface of Shell~0 in $D_0$ and $D_1$ and is 
a $C^{1-}$ extension of the world hypersurface of Shell~0 
in $D_2$, and hence $\chi$ agrees with $\chi_{(1)}$ in $D_0$ and $D_1$. 
The coordinate basis vectors are chosen so that 
they are $C^{1-}$ and agree with $\left(u_{(0)}^a, n_{(0)}^a, \thetah^a,\phih^a\right)$ defined 
as Eqs.~(\ref{u-def})--(\ref{phi-def}) at the decay event $d$. 
We use the same notation for the coordinate basis as this tetrad basis.

By using the introduced coordinate basis, the stress energy tensors of the shells are written in the form, 
\begin{align}
T_{(0)}^{ab}&=\left(\sz u_{(0)}^a u_{(0)}^b+P_{(0)}H^{ab}\right) \delta(\chi), \label{SE-0}\\
T_{(1)}^{ab}&=\left(\so u_{(1)}^a u_{(1)}^b+P_{(1)}H^{ab}\right) \delta\bigl(\chi-X_{(1)}(\tau)\bigr) 
\left|\frac{\partial \chi_{(1)}\left(\tau,\chi\right)}{\partial \chi}\right|^{-1}, \label{SE-1}\\
T_{(2)}^{ab}&=\left(\st u_{(2)}^a u_{(2)}^b+P_{(2)}H^{ab}\right) \delta\bigl(\chi-X_{(2)}(\tau)\bigr) 
\left|\frac{\partial \chi_{(2)}\left(\tau,\chi\right)}{\partial \chi}\right|^{-1}, \label{SE-2}
\end{align}
where 
$\chi=X_{(J)}(\tau)$ with $X_{(J)}(0)=0$ represents the world hypersurface of Shell~$J$ ($J=1,2$). 
We have
\begin{align}
\frac{\partial \chi_{(1)}\left(\tau,\chi\right)}{\partial \chi}
&=(d\chi_{(1)})_a\left(\frac{\partial}{\partial \chi}\right)^a, \label{del-1}\\
\frac{\partial \chi_{(2)}\left(\tau,\chi\right)}{\partial \chi}
&=\left(d\chi_{(2)}\right)_a\left(\frac{\partial}{\partial \chi}\right)^a, \label{del-2}
\end{align}
where, at the decay event $d$, 
\begin{align}
(d\chi_{(1)})_a&=\epsilon_{(1)}\sqrt{\Go^2-1}~u^{(0)}_a+\Go n^{(0)}_a, \nonumber \\
(d\chi_{(2)})_a&=\epsilon_{(2)}\sqrt{\Gt^2-1}~u^{(0)}_a+\Gt n^{(0)}_a, \nonumber
\end{align}
and 
$$
\left(\frac{\partial}{\partial\chi}\right)^a=n_{(0)}^a. 
$$
Hence we have
\begin{align}
\frac{\partial \chi_{(1)}\left(\tau,\chi\right)}{\partial \chi}
&=\Go, \\
\frac{\partial \chi_{(2)}\left(\tau,\chi\right)}{\partial \chi}
&=\Gt.  
\end{align}

\begin{figure}[htbp]
 \begin{center}
 \includegraphics[width=6cm,clip]{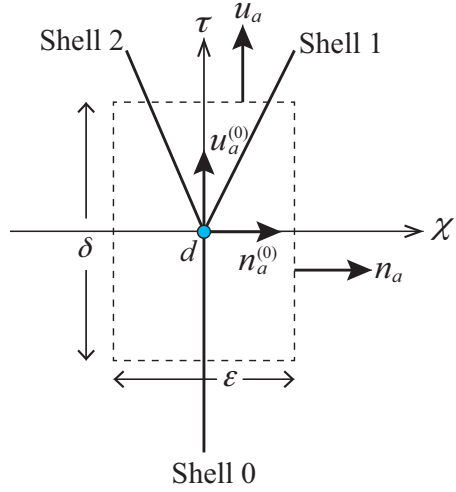}
 \end{center}
 \caption{
The domain of integration is schematically depicted by a dashed square.  }
 \label{Conservation}
\end{figure}

We integrate the Bianchi identity $\nabla_bT^{ab}=0$ over the small neighborhood of the decay event $d$ 
shown in Fig.~\ref{Conservation}: the domain of integration is chosen so that the shells do not intersect 
the boundaries $\chi=\pm \varepsilon/2$, only Shell~0 intersects the boundary $\tau=-\delta/2$, 
whereas only Shell~1 and Shell~2 intersects the boundary $\tau=+\delta/2$. Then, we have 
\begin{align}
0=&\int_{-\delta/2}^{+\delta/2}d\tau\int_{-\varepsilon/2}^{+\varepsilon/2}d\chi 
\int_{-\varepsilon/2}^{+\varepsilon/2}d\hat{\theta}\int_{-\varepsilon/2}^{+\varepsilon/2}d\hat{\phi}
\sqrt{-g}u^{(0)}_b\nabla_a T^{ab} \nonumber \\
=&\int_{-\delta/2}^{+\delta/2}d\tau\int_{-\varepsilon/2}^{+\varepsilon/2}d\chi 
\int_{-\varepsilon/2}^{+\varepsilon/2}d\hat{\theta}\int_{-\varepsilon/2}^{+\varepsilon/2}d\hat{\phi}\sqrt{-g}
\left[\nabla_a \left(T^{ab}u^{(0)}_b\right)-T^{ab}\nabla_a u^{(0)}_b\right] \nonumber \\
=&\int_{-\delta/2}^{+\delta/2}d\tau\int_{-\varepsilon/2}^{+\varepsilon/2}d\chi 
\int_{-\varepsilon/2}^{+\varepsilon/2}d\hat{\theta}\int_{-\varepsilon/2}^{+\varepsilon/2}d\hat{\phi}
\left(\sqrt{-g}T^{ab}u^{(0)}_b\right)+{\cal O}(\varepsilon^2\delta) \nonumber \\
=&\int_{-\varepsilon/2}^{+\varepsilon/2}d\chi 
\int_{-\varepsilon/2}^{+\varepsilon/2}d\hat{\theta}\int_{-\varepsilon/2}^{+\varepsilon/2}d\hat{\phi}
\left[\left.\sqrt{-g}\left(T_{(1)}^{\tau b}u^{(0)}_b+T_{(2)}^{\tau b}u^{(0)}_b\right)\right|_{\tau=+\delta/2}
-\left.\sqrt{-g}T_{(0)}^{\tau b}u^{(0)}_b \right|_{\tau=-\delta/2}\right] \nonumber \\
+&\int_{-\delta/2}^{+\delta/2}d\tau\int_{-\varepsilon/2}^{+\varepsilon/2} d\hat{\theta}
\int_{-\varepsilon/2}^{+\varepsilon/2}d\hat{\phi} 
\left(\left.\sqrt{-g}T^{\chi b}u^{(0)}_b\right|_{\chi=+\varepsilon/2}
-\left.\sqrt{-g}T^{\chi b}u^{(0)}_b \right|_{\chi=-\varepsilon/2}\right) \nonumber \\
+&\int_{-\delta/2}^{+\delta/2}d\tau\int_{-\varepsilon/2}^{+\varepsilon/2} d\chi 
\int_{-\varepsilon/2}^{+\varepsilon/2}d\hat{\phi} 
\left(\left.\sqrt{-g}T^{\hat{\theta} b}u^{(0)}_b\right|_{\hat{\theta}=+\varepsilon/2}
-\left.\sqrt{-g}T^{\hat{\theta} b}u^{(0)}_b \right|_{\hat{\theta}=-\varepsilon/2}\right) \nonumber \\
+&\int_{-\delta/2}^{+\delta/2}d\tau\int_{-\varepsilon/2}^{+\varepsilon/2} d\chi 
\int_{-\varepsilon/2}^{+\varepsilon/2}d\hat{\theta} 
\left(\left.\sqrt{-g}T^{\hat{\phi} b}u^{(0)}_b\right|_{\hat{\phi}=+\varepsilon/2}
-\left.\sqrt{-g}T^{\hat{\varphi} b}u^{(0)}_b \right|_{\hat{\phi}=-\varepsilon/2}\right) 
+{\cal O}(\varepsilon^2\delta) \nonumber \\
=&\int_{-\varepsilon/2}^{+\varepsilon/2}d\chi 
\int_{-\varepsilon/2}^{+\varepsilon/2}d\hat{\theta}\int_{-\varepsilon/2}^{+\varepsilon/2}d\hat{\phi} 
\Biggl[\left.\frac{\sqrt{-g}}{u_\tau^{(0)}}
\left(T_{(1)}^{ab}u^{(0)}_b+T_{(2)}^{ab}u^{(0)}_b\right)u_a^{(0)}\right|_{\tau=+\delta/2} \nonumber \\
&-\left.\frac{\sqrt{-g}}{u_\tau^{(0)}}T_{(0)}^{a b}u^{(0)}_a u^{(0)}_b \right|_{\tau=-\delta/2}\Biggr] 
+{\cal O}(\varepsilon^2\delta) \nonumber \\
=&\frac{\sqrt{-g}}{u_\tau^{(0)}}\left(\so\Go + \st \Gt-\sz\right)\varepsilon^2+{\cal O}(\varepsilon^2\delta), \label{MC-u2}
\end{align}
where we have used the finiteness of $\nabla_a u_b$ in the third equality, 
$\left.T^{ab}\right|_{\chi=\pm\varepsilon/2}=0$ due to the situation we consider (see Fig.~\ref{Conservation}), 
and $T^{ab}\hat{\theta}_a u_b|_{\hat{\theta}=+\varepsilon/2}
=T^{ab}\hat{\theta}_a u_b|_{\hat{\theta}=-\varepsilon/2}$
and $T^{ab}\hat{\phi}_a u_b|_{\hat{\phi}=+\varepsilon/2}=T^{ab}\hat{\phi}_a u_b|_{\hat{\phi}=-\varepsilon/2}$ 
due to the spherical symmetry 
in the forth equality, and $\left.u^{(0)}_a\right|_{\tau=\pm\delta/2}=\left.u_a^{(0)}\right|_{\tau=0}
\left[1+{\cal O}\left(\delta\right)\right]$ and Eqs.~(\ref{SE-0})--(\ref{SE-2}) 
in the final equality. Hence, in the limit of $\delta\rightarrow0$, by multiplying Eq.~(\ref{MC-u2}) by  
$4\pi\rd^2$, we obtain Eq.~(\ref{MC-u}).

By the similar manipulations to those in Eq.~(\ref{MC-u2}), we have
\begin{align}
0=&\int_{-\delta/2}^{+\delta/2}d\tau\int_{-\varepsilon/2}^{+\varepsilon/2}d\chi d\hat{\theta}d\hat{\phi}
\sqrt{-g}n^{(0)}_b\nabla_a T^{ab} \nonumber \\
=&\int_{-\varepsilon/2}^{+\varepsilon/2}d\chi d\hat{\theta}d\hat{\phi} 
\left[\left.\frac{\sqrt{-g}}{u_\tau^{(0)}}\left(T_{(1)}^{ab}n^{(0)}_b+T_{(2)}^{ab}n^{(0)}_b\right)u^{(0)}_a\right|_{\tau=+\delta/2}
-\left.\frac{\sqrt{-g}}{u_\tau^{(0)}}T_{(0)}^{ab}n^{(0)}_b u^{(0)}_a\right|_{\tau=-\delta/2}\right]
+{\cal O}(\varepsilon^2\delta) \nonumber \\
=&-\frac{\sqrt{-g}}{u_\tau^{(0)}}\left(\epsilon_{(1)}\so\sqrt{\Go^2-1} + \epsilon_{(2)} \st \sqrt{\Gt^2-1}\right)\varepsilon^2
+{\cal O}(\varepsilon^2\delta).
\end{align}
Hence, in the limit of $\delta\rightarrow 0$, by multiplying Eq.~(\ref{MC-u2}) by  
$4\pi\rd^2$, we obtain Eq.~(\ref{MC-n}).

\end{document}